\overfullrule=0pt
\baselineskip=5ex
\magnification=\magstep1
\raggedbottom
\input epsf

\font\fivepoint=cmr5
\font\sixpoint=cmr6
\font\ninepoint=cmr9
\headline={\hfill{\fivepoint QDOTS EHLJPSJY \ 3 Feb /94}}

\def\1{{\bf 1}}
\def\d{{\rm d}}
\def\A{{\bf A}}
\def\Tr{{\rm Tr}}
\def\C{{\bf C}}
\def\R{{\bf R}}
\def\E{{\cal E}}
\def\TF{{\rm TF}}
\def\MTF{{\rm MTF}}
\def\mfr#1/#2{\hbox{${{#1} \over {#2}}$}}
\def\const.{{\rm const.}}
\def\uprho{\raise1pt\hbox{$\rho$}}
 \catcode`@=11
\def\eqalignii#1{\,\vcenter{\openup1\jot \m@th
\ialign{\strut\hfil$\displaystyle{##}$& $\displaystyle{{}##}$\hfil&
$\displaystyle{{}##}$\hfil\crcr#1\crcr}}\,} \catcode`@=12
\def\boxit#1{\thinspace\hbox{\vrule\vtop{\vbox{\hrule\kern1pt
\hbox{\vphantom{\tt/}\thinspace{\tt#1}\thinspace}}\kern1pt\hrule}\vrule}
\thinspace}
\def\upchi{\raise1pt\hbox{$\chi$}}
\centerline{\bf THE GROUND STATES OF LARGE QUANTUM DOTS}
\centerline{\bf IN MAGNETIC FIELDS\footnote{$^\dagger$}{\sixpoint To appear in
Phys.~Rev.~B}}
\bigskip
\bigskip
\centerline{Elliott H. Lieb\footnote{$^*$}{\sixpoint Work partially
supported by U.S. National Science Foundation grant PHY90-19433 A03} and
Jan Philip Solovej\footnote{$^{**}$}{\sixpoint Work partially supported by
U.S. National Science Foundation grant DMS 92-03829}}
\centerline{\it Department of Mathematics, Fine Hall, Princeton University,
Princeton, NJ  08544}
\bigskip
\centerline{Jakob Yngvason\footnote{$^{***}$}{\sixpoint Work partially
supported by the Icelandic Science Foundation and the Research Fund of the
University
of Iceland.}}
\centerline{\it Science Institute, University of Iceland, Dunhaga 3, IS-107
Reykjavik, Iceland}
\bigskip
\bigskip
{\narrower\smallskip\noindent\baselineskip=0.75\baselineskip

{\bf Abstract:} The quantum mechanical ground state of a 2D $N$-electron
system in a confining potential $V(x)=Kv(x)$ ($K$ is a coupling constant)
and a homogeneous magnetic
field $B$ is studied in the high density limit $N\to\infty$, $K\to \infty$
with $K/N$ fixed. It is proved that the ground state energy and
electronic
density can be computed {\it exactly} in this limit by minimizing simple
functionals of the density. There are three such functionals depending on
the way $B/N$ varies as $N\to\infty$: A 2D Thomas-Fermi (TF) theory applies
in the case $B/N\to 0$; if $B/N\to{\rm const.}\neq 0$ the correct limit
theory is a modified $B$-dependent TF model, and the case $B/N\to\infty$ is
described by a ``classical'' continuum electrostatic theory. For
homogeneous potentials this last model describes also the weak coupling
limit $K/N\to 0$ for arbitrary $B$. Important steps in the proof are the
derivation of a new Lieb-Thirring inequality for the sum of eigenvalues
of single particle Hamiltonians in 2D with magnetic fields,
and an estimation of the exchange-correlation energy. For this last
estimate we study a model of classical point charges with electrostatic
interactions that provides a lower bound for the true quantum mechanical
energy.
\smallskip}
\bigskip \bigskip

\vfil\eject
\bigskip\noindent
{\bf I INTRODUCTION} \medskip

In the last few years considerable experimental and theoretical work has
been devoted to the study of quantum dots, which are atomic-like
two-dimensional systems, confined within semiconductor heterostructures.
\footnote{*}{\ninepoint  The number of articles on this subject is by
now quite large.
See, e.g.,  [1], [2] for
reviews, [3]-[7] for recent measurements of conductivity and
capacity of quantum dots, and [8]-[18] for various theoretical aspects
and further references.}
The parameters of
such artificial atoms may differ appreciately from their natural
counterparts because of the interactions of the electrons with the crystal
where they reside. In a quantum dot the natural atomic unit of length is
$a_*=\epsilon\hbar^2/(m_*e^2)$, where $\epsilon$ is the dielectric constant
and $m_*$ is the effective electron mass. Compared with the usual Bohr
radius, $a_0=\hbar^2/(me^2)$, the length $a_*$ is typically large, e.g.,
$a_*\approx 185\ a_0$ in GaAs. The corresponding natural unit, $B_*$, with
which we measure the magnetic field, $B$, is the field at which the
magnetic length $\ell_B=\hbar e/(B^{1/2}c)$ equals $a_*$, i.e.,
$B_*=(a_0/a_*)^2B_0$, where $B_0=e^3m^2c/\hbar^3=2.35\times 10^5$ T is the
value corresponding to free electrons. If $a_0/a_*$ is small, $B_*$ can be
much smaller than $B_0$. Thus $B_*\approx 7$ T in GaAs. This makes it
possible to study in the laboratory effects which, for natural atoms,
require the magnetic fields of white dwarfs or even neutron stars.

The ground state properties of natural atoms in high magnetic fields have
recently been analyzed rigorously in the asymptotic limit where the number
of electrons and the nuclear charge are large [19-21]. For artificial atoms
one may expect asymptotic analysis to be even more useful because the accuracy
increases with the number of electrons, and a quantum dot can easily
accommodate several hundred or even a thousand electrons. In the present
paper we carry out such an analysis of the ground state of a quantum dot in
a magnetic field. One of our conclusions is that the \lq\lq self
consistent model\rq\rq \ introduced by McEuen {\it et al} [3,13] is a
rigorous limit of quantum mechanics. This model has recently been
applied to explain interesting features of the addition
spectra of large quantum dots in strong magnetic fields [15,6].

Before discussing our results for dots we summarize, for comparison,
the main findings about atoms in [19-21] . The quantum mechanical ground
state energy and electronic density of a natural atom or ion with electron
number $N$ and nuclear charge $Z$ in a homogeneous magnetic field $B$ can,
in the limit
$N\to\infty$, $Z\to\infty$ with $Z/N$ fixed, be described exactly by
functionals of the density, or, in one case, of density matrices. There are
five different functionals, depending on the way $B$ varies with $N$ as
$N\to \infty$. In each of the cases $B\ll N^{4/3}$ (with $B$ measured in
the natural unit $B_0$), $B\sim N^{4/3}$ and $N^{4/3}\ll B\ll N^3$ the
correct asymptotics is given by an appropriate functional of the
semiclassical Thomas-Fermi type. For $B\sim N^3$ a novel type of
functional, depending on density matrices is required, whereas the case
$B\gg N^3$ is described by a density functional that can be minimized in
closed form.

A review of our results about quantum dots was given in [22].
Due to the reduced dimensionality of the electronic motion, there are only
three different asymptotic theories for quantum dots instead of five for
natural atoms. These three theories are given by simple functionals of the
density and correspond respectively to the cases $B\ll N$, $B\sim N$ and
$B\gg N$ ($B$ measured in units of $B_*$) as $N\to\infty$ with $V/N$ fixed,
where $V$ is the attractive exterior potential
that restricts the two-dimensional motion of the electrons. This potential,
which plays the same role as the nuclear attraction in a natural atom, is
generated in a quantum dot by exterior gates, and thus is adjustable to a
certain extent. In the course of proving the asymptotic limits we shall
also consider, in addition to the density functionals, a model of classical
point charges in two dimension that gives a lower bound to the
quantum-mechanical
energy.

Some of the methods and results of the present paper contrast markedly with
those of our earlier work [19-21]. From a mathematical point of view the
most interesting feature of quantum dots compared to natural atoms is the
somewhat peculiar electrostatics that appears because the interaction
between the electrons is given by the three dimensional Coulomb potential
although the motion is two dimensional. Also, the fact that the kinetic
energy  vanishes in the lowest Landau level requires additional mathematical
effort in order to bound the kinetic energy from below by a functional of
the density.
We now describe in
more detail the limit theorems to be proved in the sequel. A quantum dot
with $N$
electrons in a confining potential $V$
and a homogeneous magnetic field $B$ is modeled by the following
Hamiltonian:
$$H_N = \sum \limits^N_{j=1} H^{(j)}_1 + {e^2\over \epsilon} \sum
\limits_{1 \leq i < j \leq N} \vert x_i - x_j \vert^{-1}, \eqno(1.1)$$
with
$x_i \in \R^2$ and where $H_1$ is the one-body Hamiltonian
$$H_1 = {\hbar^2
\over 2m_*} \left( i \nabla - {e \over \hbar c} {\bf A} \right)^2 + g_*
\left( {\hbar e \over 2mc} \right) {\bf S}\cdot {\bf B} -
\left( {\hbar e \over 2m c}
\right) \left({m\over m_*} - {\vert g_*\vert \over 2} \right) B + V (x).
\eqno(1.2)$$
As before, $e$ and $m$ denote the charge and mass of a (free)
electron, $\epsilon$ is the dielectric constant, $m_*$ is the effective mass,
and $g_*$ is the effective $g$-factor. The magnetic vector potential is ${\bf
A}(x) = \mfr1/2 (-Bx^1, Bx^2)$ (with $x=(x^1,x^2)\in\R^2$), ${\bf
B}=(0,0,B)$ and ${\bf S}$ is the vector of electron spin operators.
The potential $V(x)$ is supposed
to be continuous and confining, which is to say that $V(x) \rightarrow
\infty$ as $\vert x \vert \rightarrow \infty$. It is not assumed to be
circularly symmetric. The constant term in (1.2),
$-(\hbar e/(2mc)) ((m/m_*)-\vert g_*\vert/2) B$, is included in order that
the ``kinetic energy'' operator, $H_{\rm kin}=H_1 - V (x)$, has a
spectrum starting at zero. The Hilbert space is that appropriate for
fermions with spin, the antisymmetric tensor product $\bigwedge \limits^N_1
L^2
(\R^2; \C^2)$.

We define an effective charge by $e_*=e/\sqrt\epsilon$ and choose units
such that $\hbar=m_*=e_*=1$. The unit of length is then the effective Bohr
radius $a_* = \hbar^2/(m_* e^2_*)$ and the unit of energy is $E_* = e_*^2/a_*=
e^4_* m_*/\hbar^2.$ Moreover, the unit $B_*$ for the magnetic field is
determined
by $\hbar eB_*/(m_*c)=E_*$, so $B_* = e^3_* m^2_*
c/(\epsilon^{1/2}\hbar^3).$ The values for GaAs are $a_*=9.8$ nm, $E_*=12$
meV and $B_*=6.7$ T.

The true quantum-mechanical ground state energy of $H_N$ is denoted by
$E^{\rm Q}(N,B,V)$ and the true ground state electron density by $\rho^{\rm
Q}_{N,B,V}(x)$. The density functionals that describe the asymptotics of
$E^{\rm Q}$ and $\rho^{\rm Q}$ are of three types. The first is a standard
two-dimensional Thomas-Fermi energy functional
$$\E^{\TF}[\rho;V]=(\pi/2)\int\rho(x)^2\d x+\int V(x)\rho(x)\d
x+D(\rho,\rho)\eqno(1.3)$$ with $$D(\rho,\rho)={1\over
2}\int\int{\rho(x)\rho(y)\over\vert x-y\vert}\d x\d y.\eqno(1.4)$$ Here
$\rho$ is a nonnegative density on $\R^2$ and all integrals are over
${\bf R}^2$ unless otherwise stated. The second functional is a
two-dimensional \lq\lq magnetic Thomas Fermi functional\rq\rq\
$$\E^{\MTF}[\rho;B,V]=\int j_B(\rho(x))\d x+\int V(x)\rho(x)\d
x+D(\rho,\rho)\eqno(1.5)$$ where $j_B$ is a piece-wise linear function that
will be defined precisely in the next section. This functional is the two
dimensional analogue of the three dimensional magnetic Thomas Fermi
functional that was introduced in [23] and further studied in [24,25,21].
The present two dimensional version was first stated in [3]; these authors
call it the self-consistent model (SC)\footnote{$^*$}{\ninepoint The repulsion
term
considered in [3] is slightly different from
$D(\rho,\rho)$, since it has cut-offs at long and short distances. It is
still positive definite as a kernel and our methods can easily be adapted to
prove Theorems 1.1 and 1.2 with such cut-off Coulomb kernels. }.

The last asymptotic functional will be called the \lq\lq classical
functional\rq\rq, since the kinetic energy term is absent and only
classical interactions remain:
$$\E^{\rm C}[\rho;V]=\int
V(x)\rho(x)\d x+D(\rho,\rho).\eqno(1.6)$$

The functionals (1.3) and (1.6) are in fact limiting cases of (1.5)
for $B\to 0$ and $B\to\infty$ respectively.
As discussed in detail later, for each functional there is a unique
density that minimizes it under the constraint $\int\rho=N$.
We denote these densities respectively by $\rho^{\TF}_{N,V}(x)$,
$\rho^{\MTF}_{N,B,V}(x)$, $\rho^{\rm C}_{N,V}(x)$, and the corresponding
minimal energies by $E^\TF(N,V)$, $E^\MTF(N,B,V)$ and $E^{\rm C}(N,V)$.

In order to relate $E^{\rm Q}$ to these other
energies we take a high density limit. This is achieved by  letting $N$ tend
to infinity (which is a reasonable thing to do physically, since $N$ can be
several hundred) and we let $V$ tend to infinity. The latter statement
means
that we fix a potential $v$ and set $V=Nv$. With this understanding of
$N,V\to\infty$
our main results are summarized in the following two theorems.
[In order to prove these theorems we need to assume that $V$ is
sufficiently
regular. The technical requirement is that $V$ belongs to the class
$C^{1,\alpha}_{\rm loc}$,
(see Theorem~3.2 for the definition of $C^{1,\alpha}_{\rm loc}$)].

{\it 1.1 THEOREM (Limit theorem for the energy).}
\nobreak
\noindent{\it Let $V=Nv$ with $v$ a fixed function in $C^{1,\alpha}_{\rm loc}$.
Then
$$\lim_{N\to\infty} E^{\rm Q}(N,B,V)/E^\MTF(N,B,V)=1\eqno(1.7)$$
uniformly in $B$. Moreover,
$$\lim_{N\to\infty} E^{\rm Q}(N,B,V)/E^\TF(N,V)=1\qquad \hbox{if
}B/N\to 0\eqno(1.8)$$
and}
$$\lim_{N\to\infty} E^{\rm Q}(N,B,V)/E^{\rm C}(N,V)
=1\qquad \hbox{if
}B/N\to \infty.\eqno(1.9)$$

{\it 1.2 THEOREM (Limit theorem for the density).}
\nobreak
\noindent
{\it Let $V=Nv$ with $v$ a fixed function in $C^{1,\alpha}_{\rm loc}$.
Then
$${1\over N}\rho^{\rm Q}_{N,B,V}\to\rho^{\MTF}_{1,B/N,v}\eqno(1.10)$$
uniformly in $B$, and
$${1\over N}\rho^{\rm Q}_{N,B,V}\to\rho^{\TF}_{1,v}\qquad \hbox{if }
B/N\to
0,\eqno(1.11)$$
$${1\over N}\rho^{\rm Q}_{N,B,V}\to\rho^{\rm C}_{1,v}\qquad \hbox{if }
B/N\to
\infty.\eqno(1.12)$$
The convergence is in the weak $L^1$ sense\footnote{$^*$}{\ninepoint By
definition, a sequence of functions $f_n$ converges to a function $f$ in
weak $L^1$ sense if $\int f_ng\to\int fg$ for all bounded (measurable)
functions $g$.}} 

Let us add a few comments on these results. As discussed in
the Section II,
the energy $E^\MTF$ has the
scaling property
$$E^\MTF(N,B,V)=N^2E^\MTF(1,B/N,V/N).\eqno(1.13)$$
Thus (1.7) is equivalent to
$$E^{\rm Q}(N,V,B)=N^2E^\MTF(1,B/N,V/N)+o(N^2)\eqno(1.14)$$
where the error term is uniformly bounded in $B$ for $V/N$ fixed.
One expects the error to be $O(N^{3/2})$, which is
the order of the exchange contribution to the Coulomb interaction, but
our methods do not quite allow us to prove this. We do,
however, show that for $B/N$
larger than a critical value (depending on $V/N$) one has
$E^\MTF(N,B,V)=E^{\rm C}(N,V)$ and
$$E^{\rm Q}(N,V,B)\geq N^2E^{\rm C}(1,V/N)-bN^{3/2}\eqno(1.15)$$
where the coefficient $b$ depends only on $V/N$.

The condition that $V/N$ is fixed as
$N\to\infty$ guarantees that the diameter of the electronic density
distribution stays
bounded as $N\to\infty$; thus the limit we are considering is really
a high density limit rather than simply a large
$N$ limit. On the other hand, for a homogeneous potential $V$
(e.g., quadratic, as is often
assumed) one obtains also
a nontrivial $N\to\infty$ limit for  $V$ {\it fixed\/}, if the lengths are
suitably scaled. In fact, this limit is given by the classical functional
(1.6).
Intuitively this is easy to understand, for if an increase in $N$ is not
compensated by an increase in
$V$ the charge density spreads out and the kinetic energy terms in (1.5)
and
(1.3) become negligible compared with the other terms. (The result again
requires $V$ to be in $C^{1,\alpha}_{\rm loc}$.)

{\it 1.3 THEOREM (Energy limit with a homogeneous potential).}
\nobreak
\noindent{\it Assume that $v$ is homogeneous of degree $s\geq1$, i.e.,
$$
	v(\lambda x)=\lambda^sv(x).
$$
Then
$$\lim_{N\to\infty} E^{\rm Q}(N,B,Kv)/E^\MTF(N,B,Kv)=1\eqno(1.16)$$
uniformly in $B$ and in $K$ as long as $K/N$ is bounded above.
Moreover, if $K/N\to 0$ as $N\to\infty$, then
$$\lim_{N\to\infty} E^{\rm Q}(N,B,Kv)/E^{\rm C}(N,B)=1\eqno(1.17)$$
uniformly in $B$.}

One can also prove a limit theorem for the density in the case of
homogeneous
potentials. Since the formulation of such a theorem becomes somewhat
complicated
we refrain from doing this, but refer to Eqs. (2.14)-(2.16)  below for
the scaling of the MTF functional with $k=K/N$ and to (3.24) for the weak
coupling limit of the MTF density.

The proof of the limit theorems involves the following steps. In sections
II and III we discuss the basic properties of the functionals (1.3)-(1.6).
In Section IV we consider the energy of a system of classical point charged
particles in $\R^2$ in the exterior potential $V$ as a function of the
positions of
the charges. This energy has a minimum, denoted by $E^{\rm P}(N,V)$
(with ``P'' denoting ``particle'').
A significant remark is that the
charge configuration, for which the minimum is obtained, is confined within
a radius independent of the total charge $N$ for fixed
$V/N$. This ``finite radius
lemma'', which also holds for the charge densities minimizing the
functionals (1.3)-(1.6), is proved in an appendix. Using this and an
electrostatics
lemma of Lieb and Yau [26] we derive the bounds
$$E^{\rm C}(N,V)-aN^{3/2}\geq E^{\rm P}(N,V)\geq E^{\rm
C}(N,V)-bN^{3/2}\eqno(1.18)$$
where $a$ and $b$ depend only on $V/N$. These bounds are of
independent interest apart from their role in the proof of the limit
theorems where, in fact, only the latter inequality is needed. Upper and
lower bounds to the
quantum mechanical energy $E^{\rm Q}(N,B,V)$ in terms of $E^\MTF(N,B,V)$
with controlled errors are derived in the final Section V. The upper bound
is a
straightforward variational calculation using magnetic coherent states in
the same way as
in [21]. For the lower bound one treats the cases of large $B$ and small
$B$ separately. The estimate for large $B$ is obtained by first noting
that obviously $E^{\rm Q}(N,B,V)\geq E^{\rm P}(N,V)$, because the kinetic
energy is nonnegative,  and then using
(1.18). For small $B$ two auxiliary results are required: A generalization
of the magnetic Lieb-Thirring equality considered in [21],
and an estimate of the correlation
energy. Once these have been established the proof of Theorem 1.1 is
completed by
coherent state analysis. The limit theorem for the density follows easily
from the limit theorem for the energy by perturbing $V$ with bounded
functions.

\bigskip
\vbox{\noindent
{\bf II. THE MTF THEORY: ITS DEFINITION AND PROPERTIES}

By employing the natural units defined in the introduction, the \lq\lq kinetic
energy\rq\rq\ operator can be written}
$$H_{\rm kin}=\mfr 1/2 \left( i \nabla - {\bf A} \right)^2 +
\gamma {\bf S}\cdot {\bf B} -
\mfr 1/2\left(1-\vert \gamma\vert
\right)  B\eqno(2.1)$$
with $\gamma=g_*m_*/(2m)$.
The spectrum of $H_{\rm kin}$ is
$$\varepsilon_{n,\sigma}=\left(n+\gamma \sigma+\mfr1/2\vert
\gamma\vert\right)B\eqno(2.2)$$
with $n=0,1,\dots$, $\sigma=\pm 1/2$.
We write the energy levels (2.2) in {\it strictly} increasing order as
$\varepsilon_\nu(B)$, $\nu=0,1,\dots$.
The degeneracy of each level per unit area is $d_\nu(B)=B/(2\pi)$, except
if, by coincidence, $\gamma$ happens to be an integer; in that case
$d_\nu(B)=B/(2\pi)$ for $\nu=0,\dots,\vert \gamma\vert-1$, while
$d_\nu(B)$ is twice as large for
the higher levels.
It is worth recalling that if $V(x) = K\vert x\vert^2$, the spectrum of the
one-body Hamiltonian $H_1$ in (1.2) is solvable.
The spectrum of $H_1$ was determined by Fock [27] in 1928, two
years before Landau's paper on the spectrum of $(i\nabla-{\bf A})^2$.
For the Hamiltonian without spin, namely
${1\over 2}(i \nabla -{\bf A})^2 + K \vert x\vert^2$, the spectrum is given by
$$E = {1\over 2}(n_1 - n_2) B + {1\over 2}(n_1 + n_2 + 1) [4K + B^2]^{1/2}
$$
with $n_1, n_2 = 0, 1, 2, \dots$.
It is remarkable that this simple spectrum gives a qualitatively
good fit to some of the data [8].

For a gas of noninteracting fermions with the energy spectrum (2.2)
the energy density $j_B$ as a function of the particle density
$\uprho$ is given by
$$\eqalign{j_B(0)&=0\cr
j_B^\prime(\uprho)&=\varepsilon_\nu
(B)\qquad\hbox{if}\quad D_{\nu}(B)<\uprho< D_{\nu+1}(B), \quad
\nu=0,1\dots\cr}$$
where $j_B^\prime=dj_B/d\rho$ and
$$D_{\nu}(B)=\sum_{\nu'=0}^\nu d_{\nu'}(B).$$
More explicitly,
$$j_B(\uprho)=\sum_{\nu=0}^{\nu_{\rm max}}\varepsilon_\nu(B)d_\nu(B)
+(\uprho-D_{\nu_{\rm max}}(B))\varepsilon_{\nu_{\rm
max}+1}(B)\eqno(2.3)$$
where $\nu_{\rm max}=\nu_{\rm max}(\uprho,B)$ is defined by
$$D_{\nu_{\rm max}}(B)\leq\uprho<D_{\nu_{\rm max}+1}(B).$$
Thus $j_B$ is a convex, piece-wise linear function with
$j_B(\uprho)=0$ for $0\leq \uprho\leq d_1(B)$. It has the scaling
property
$$j_B(\uprho)=B^2j_1(\uprho/B).\eqno(2.4)$$
As $B\to 0$,  $j_B$ becomes a quadratic function of the density:
$$\lim_{B\to 0}j_B(\uprho)=j_0(\rho)={\pi\over 2}
\uprho^2.\eqno(2.5)$$
Moreover,
$$j_B(\uprho)\leq j_0(\uprho)\eqno(2.6)$$
for all $\uprho$ and $B$ (see Fig. 1).
\medskip
\medskip
\hbox to\hsize{\hss\vbox{\hsize=0.6\hsize\epsfxsize=0.8\hsize
\hbox to\hsize{\hss\epsfbox{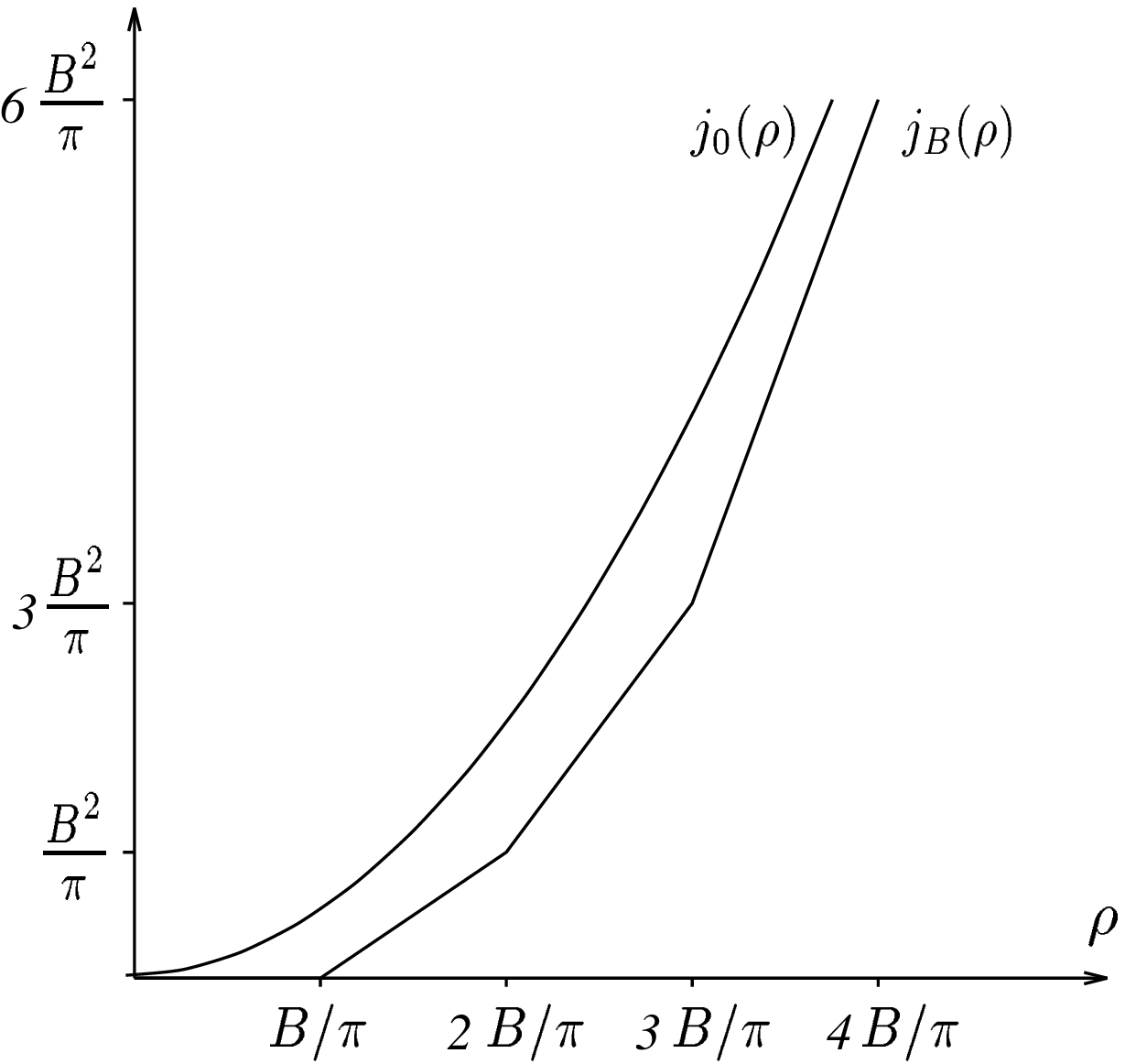}\hss}
\noindent {\bf Fig. 1} The \lq\lq kinetic\rq\rq energy densities $j_B(\rho)$
and $j_0(\rho)$ in the special case where $\gamma=0$}\hss}
\medskip

Given an exterior potential $V$ the MTF functional is defined by (1.5).
We assume that $V$ is continuous\footnote{*}{\ninepoint $V$ measurable and
locally bounded would suffice} and tends to $\infty$ as
$\vert x\vert\to\infty$. In particular, $V$ is bounded below and, by
adding a constant if necessary, we may assume that $V(x)\geq 0$ everywhere.
Because of (2.6) the functional (1.5) is defined for all nonnegative
functions $\rho$ such that
$\int \rho V<\infty$,
$\int\rho^2<\infty$ and $D(\rho,\rho)<\infty$. Since $V\geq 0$
the functional is nonnegative.
If $N$ is some positive number we denote
$${\cal C}_N=\{\rho\mid \rho\geq 0,\ \hbox{$\int$}\rho V<\infty,\
\hbox{$\int$}\rho^2<\infty,\ D(\rho,\rho)<\infty,\ \hbox{$\int$} \rho=N\}$$
and define the MTF energy  by
$$E^\MTF(N,B,V)=\inf_{\rho\in{\cal C}_N}{\cal E}[\rho;B,V].\eqno(2.7)$$
Because of (2.4) the energy satisfies the scaling relation
$$E^\MTF(N,B,V)=N^2E^\MTF(1,B/N,V/N).\eqno(2.8)$$

In the limit $B\to 0$ the kinetic energy density (2.3) converges to
$j_0(\rho)=
(\pi/2)\uprho^2$ and (1.5) converges to the energy functional (1.3) of
two-dimensional TF theory at $B=0$. It is easy to see that also
$\lim_{B\to 0}E^\MTF(N,B,V)=E^\TF(N,V)$, where $E^\TF$ is defined in the
same way as $E^\MTF$ with (1.3) replacing (1.5).
We can thus consider the TF theory as
a special case of MTF theory.
In the opposite limit, $B\to \infty$,
the kinetic energy term vanishes altogether and one obtains a classical
electrostatic model (1.6) that we shall study in Section III. Note also
that
since
$j_B\leq j_0$ for all $B$ it follows that $E^\MTF(N,B,V)\leq E^\TF(N,V)$
for all
$B$.
In particular $E^\MTF(1,\beta,v)$ is
uniformly bounded in the parameter $\beta=B/N$ for fixed $v=V/N$.

For fixed $B$ and $V$, $E^\MTF(N,B,V)$ is a convex, continuously
differentiable
function of $N$ and, since $V\geq 0$, it is monotonically increasing.
By the methods of [28], [29], [21] (see also
[30]) it is straightforward to prove the existence and uniqueness of a
minimizer
for the variational problem (2.7).

{\it 2.1 THEOREM (Minimizer).} {\it There is a unique density
$\uprho^\MTF_{N,B,V}\in{\cal C}_N$
such that $$E^\MTF(N,B,V)=\E^\MTF(\uprho^\MTF_{N,B,V}).$$}
Note that the existence of a minimizing density with $\int \uprho=N$ is
guaranteed for all $N$ because $V(x)\to\infty$ as $\vert x\vert\to\infty$.
This condition on $V$  also implies that
$\uprho_{N,B,V}$ vanishes outside a ball of finite radius, cf. Lemma~A1 in
the Appendix. The scaling relation for the minimizing density is
$$\uprho^\MTF_{N,B,V}(x)=N\uprho^\MTF_{1,B/N,V/N}(x).\eqno(2.9)$$
Theorem~2.1 includes the TF theory as a special case. In the same way as
in Prop. 4.14 in [21] one shows that
$\uprho^\MTF_{N,B,V}\to\uprho^\TF_{N,V}$ weakly in $L^1$ as $B\to 0$.

The shape of the electronic density in the
case of a quadratic potential $V(x)=K\vert x\vert^2$ and $\gamma=0$ is
shown in Fig.\ 2 for different values of $B$.  The pictures were
prepared by Kristinn Johnsen.  At the highest value of $B$ (8T), the
density is everywhere below $d_0(B)$ and given by the minimizer (3.15)
of the classical functional (1.6). At $B=$7T, all the electrons are
still in the lowest Landau level, but that level is full around the
middle of the dot where the density is anchored at $d_0(B)$. As the
field gets weaker it becomes energetically favorable for electrons at
the boundary of the dot, where the potential is high, to move into the
next Landau level close to the minimum of the potential. A dome-shaped
region then arises above the plateau at $\rho=d_0(B)=D_0(B)$, but
eventually the density hits the next plateau at $\rho=D_1(B)$. This
gradual filling of Levels continues as the field strength goes down. At
$B=2$T three Landau levels are full and electrons in the central dome
are beginning to occupy the fourth level. Finally, at $B=0$, we have
the usual Thomas-Fermi model, which may be regarded as a limiting case
with infinitely many Landau levels occupied.

\vskip 1cm
\hbox{
\epsfxsize=0.5\hsize
\epsfysize= 0.34\hsize
\epsfbox{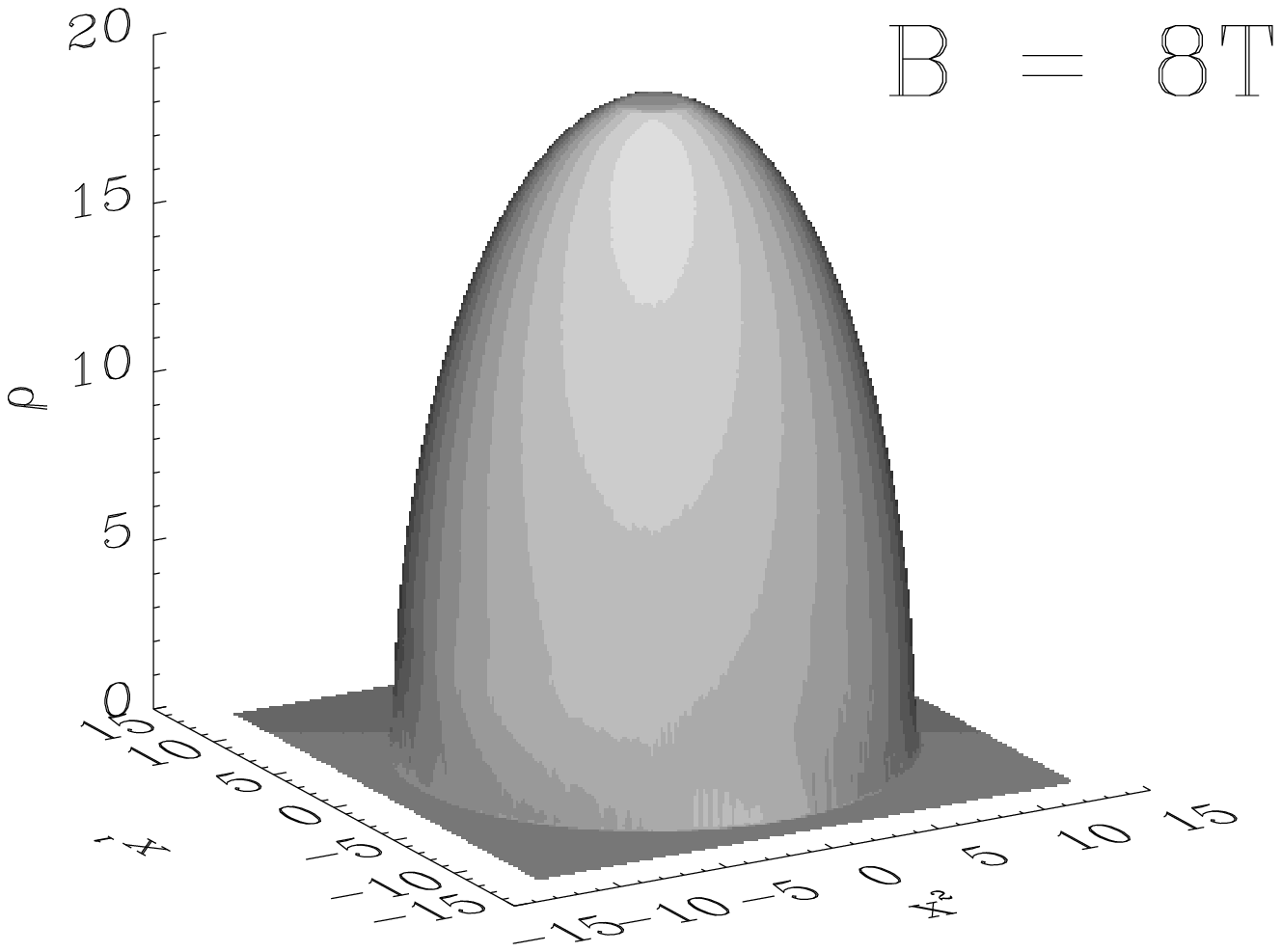}
\epsfxsize= 0.5\hsize
\epsfysize= 0.34\hsize
\epsfbox{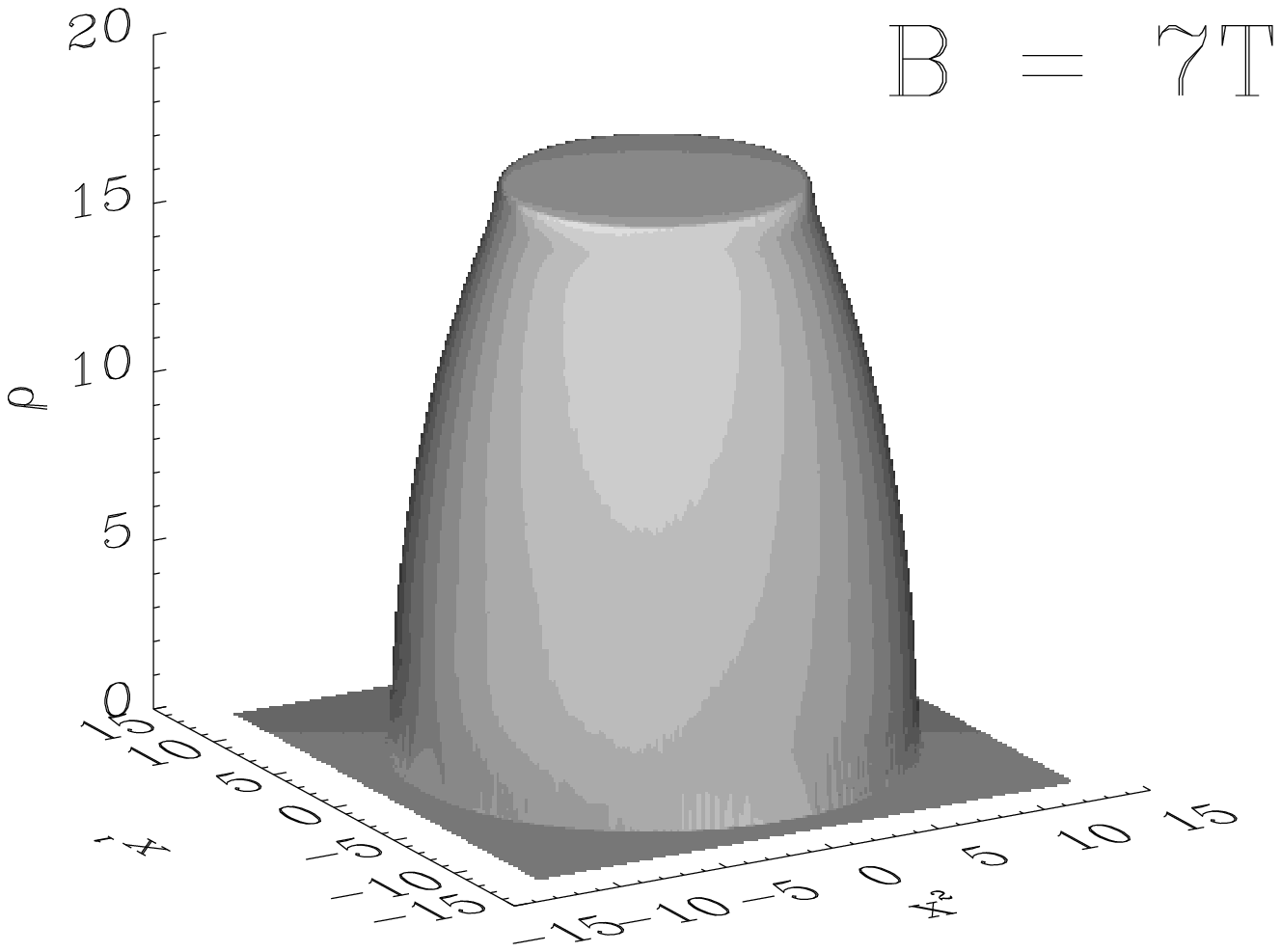}
}
\hbox{
\epsfxsize= 0.5\hsize
\epsfysize= 0.34\hsize
\epsfbox{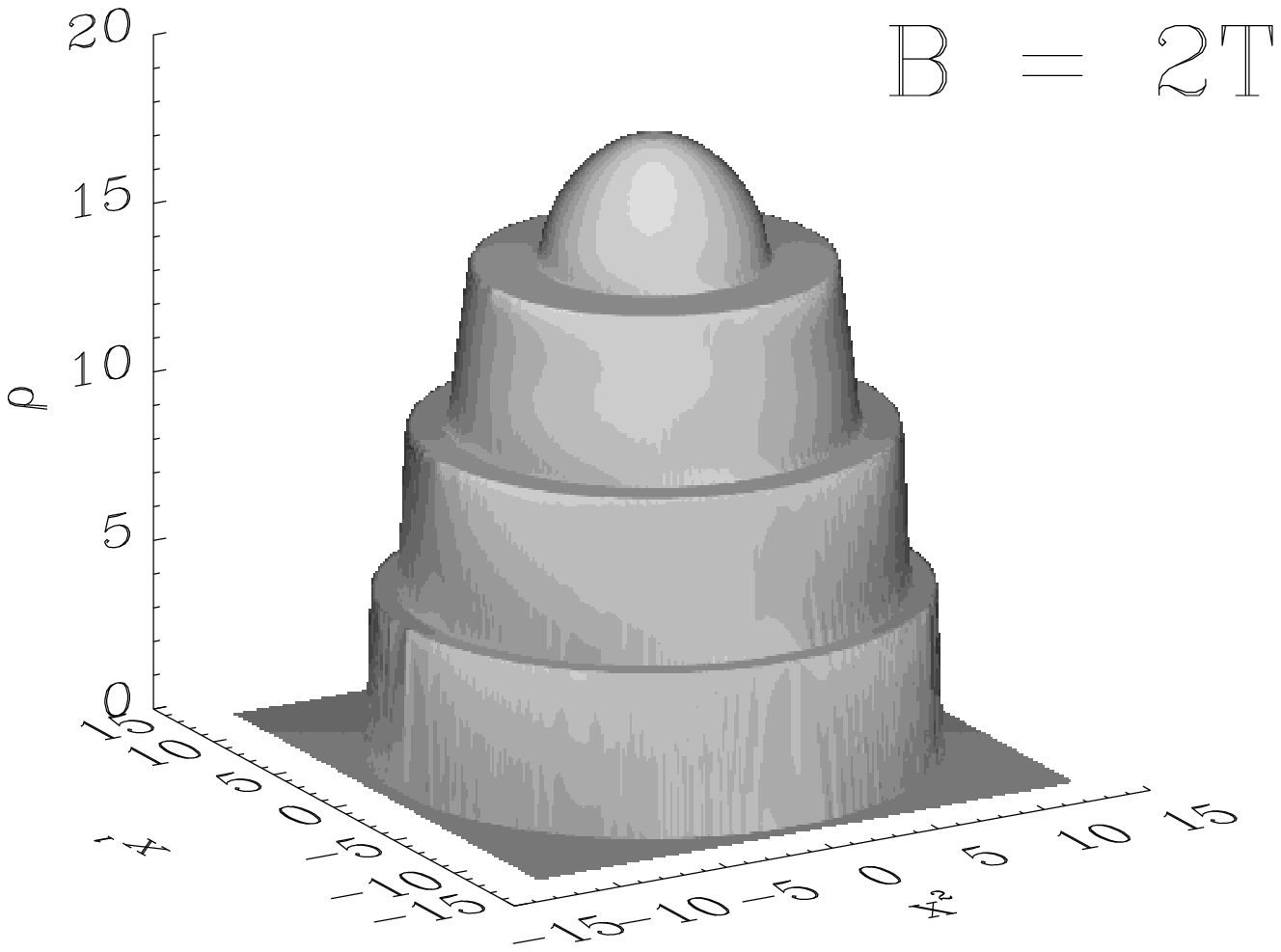}
\epsfxsize= 0.5\hsize
\epsfysize= 0.34\hsize
\epsfbox{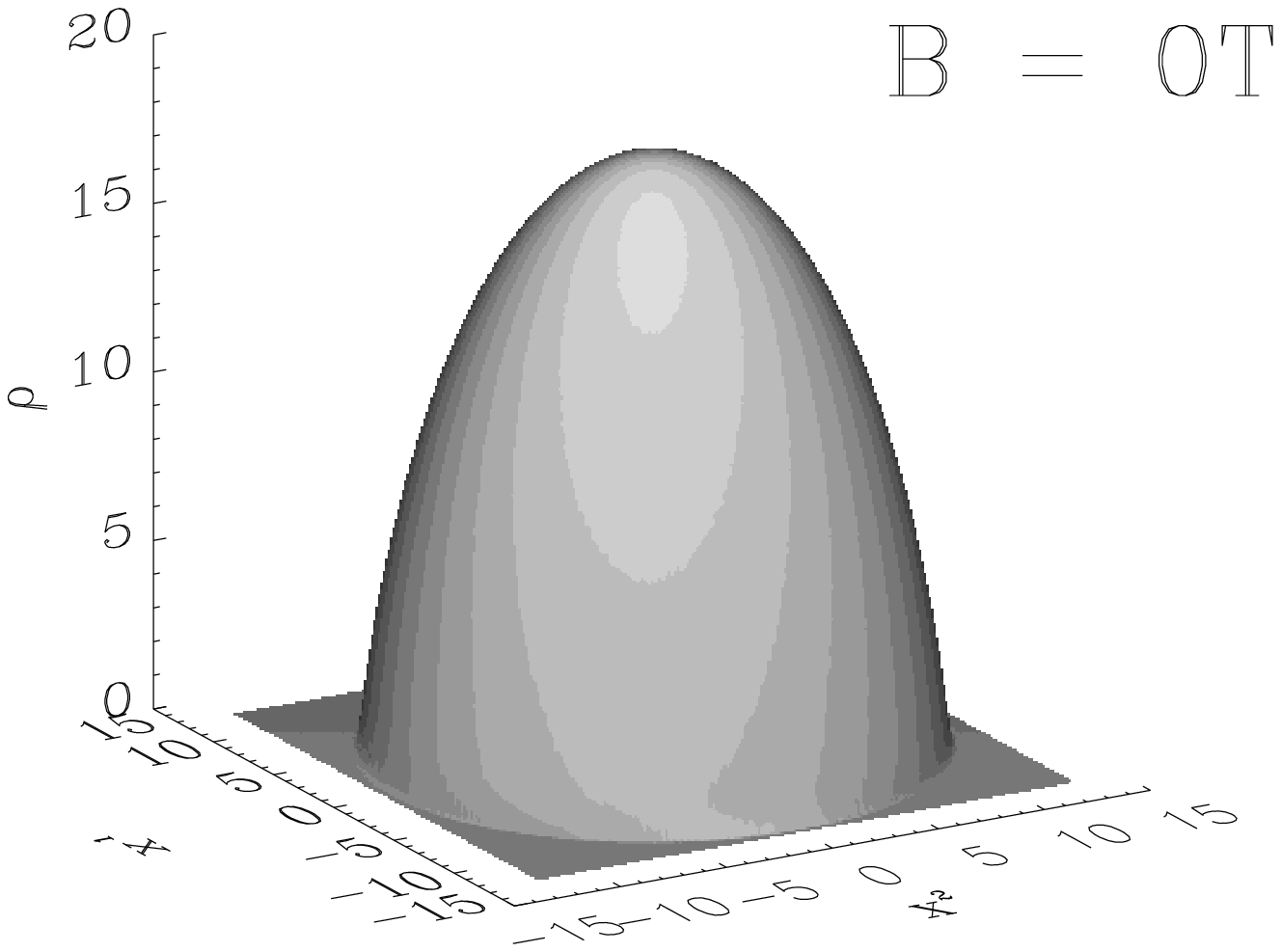}
}
\smallskip
\vbox{\noindent {\bf Fig.\ 2} Quantum dots at various magnetic field strengths.
The potential is
$V(x)=(1/2)m_*\omega^2 \vert x\vert^2$, with $m_*=0.67 m$, $\hbar\omega=3.37$
meV, and $N=50$. The coordinate
axes are displayed in units of $10^{-8}$ m and the density $\rho$ in
the units $10^{-14}{\rm  m}^{-2}$.}
\medskip
\medskip

In order to state the variational equation for the minimization problem
it is convenient to define the
derivative $j'_B=dj_B/d\rho$ of the kinetic energy density everywhere,
including points of discontinuity, as a {\it set valued} function (cf.
[30]), namely
$$j'_B(\uprho)=\cases{
\{\varepsilon_\nu(B)\}&for $D_{\nu}(B)<\uprho<D_{\nu+1}(B)\ ,\
\nu=0,1,\dots$\cr
[\varepsilon_{\nu}(B),\varepsilon_{\nu+1}(B)]&for
$\uprho=D_{\nu+1}(B),\
\nu=0,1,\dots$\cr}\eqno(2.10)$$
With this notation the Thomas-Fermi equation for the functional (1.5)
may be written as follows

{\it 2.2 THEOREM (Thomas-Fermi equation).} {\it There is a nonnegative
number
$\mu=\mu(N,B,V)$  such that the minimizer
$\uprho=\uprho^\MTF_{N,B,V}$
satisfies
$$\mu-V(x)-\rho*\vert x\vert^{-1}\cases{\in j'_B(\uprho(x))&if
$\rho(x)>0$\cr
\leq 0&if $\rho(x)=0$\cr}.\eqno(2.11)$$}
The quantity $\mu$ appearing in the TF equation is the physical chemical
potential, i.e.,
$$\mu=\partial E(N,B,V)/\partial N.\eqno(2.12)$$
Since $E$ is convex as a function of $N$, $\mu$ is monotonically
increasing with $N$ for fixed $B$ and $V$. It satisfies
$$\mu(N,B,V)=N\mu(1,B/N,V/N).\eqno(2.13)$$

The derivation of the TF equation is analogous to that in [29].
It is also true that if $(\uprho,\mu)$ is {\it any}
solution pair for (2.11), then
$\uprho$ is the minimizer of $\E$ for some $N$ and $\mu=\mu(N,B,V)$. The
proof of this is a bit trickier than in the standard case [29], because
$j_B$ is not continuously
differentiable. It has been carried out by Lieb and Loss [30].

Finally we discuss the relationship between the
MTF theory and the classical theory defined by the functional (1.6).
We of course have that
$$
	\E^{\MTF}[\rho;B,V]=\int j_B(\uprho(x))\d x +\E^{\rm C}[\rho;V].
$$
{F}rom the definition of $j_B$ one expects that the kinetic
energy term above can be neglected for large $B$ and
hence that $\lim_{B\to\infty}E^{\MTF}=E^{\rm C}$.
The rigorous proof of this fact relies on a careful study of the classical
problem. This analysis is far from trivial and is postponed to the next
section.

There is another case where the
MTF energy can be related to the classical energy.
Namely, for a homogeneous exterior potential, i.e.,
$$V(\lambda x)=\lambda^s V(x)$$
for all $\lambda>0$ with some $s>0$. We consider the potentials
$kV(x)$ with $k>0$ and are interested in the dependence of the MTF
energy and density on the coupling constant $k$.

Writing
$$\uprho(x)=k^{2/(s+1)}\hat\rho(k^{1/(s+1)}x)\eqno(2.14)$$
we have the scaling
$$\eqalignno{\E^{\MTF}[\uprho;B,kV]&=
	k^{2/(s+1)}\int j_{b}(\hat\rho)
	+\E^{\rm C}[\rho;kV]\cr
	&=k^{1/(s+1)}\left[k^{1/(s+1)}\int j_{b}(\hat\rho)
	+\E^{\rm C}[\hat\rho;V]\right],&(2.15)}$$
where
$$b=B k^{-2/(s+1)}.\eqno(2.16)$$
Changing  $k$ is thus equivalent to changing the kinetic energy by a
multiplicative factor and rescaling the magnetic field, keeping the
potential fixed.
We shall show in the next section that for $k$ small  $E^{\rm C}$
is a good approximation to $E^{\MTF}$.

\bigskip
\vbox{\noindent
{\bf III.  THE CLASSICAL CONTINUOUS MODEL:  A LIMIT OF MTF THEORY}

For densities $\uprho$ small enough $(\uprho (x) \leq d_1 (B)$ for all $x$)
the kinetic energy $j_B (\uprho)$ vanishes. It is therefore natural to
consider the resulting {\bf classical energy functional} defined by
(1.6), i.e.,}
$$\E^{\rm C} [\uprho ; V] = \int \uprho (x) V(x) \d x + \mfr1/2 \int\int
\uprho (x) \vert x-y \vert^{-1} \uprho (y) \d x \d y. \eqno(3.1)$$ The
corresponding {\bf classical energy} is $$E^{\rm C} (N,V) = \inf \left\{
\E^{\rm C} [\uprho ; V]: \uprho \geq 0, \int \uprho = N
\right\}.\eqno(3.2)$$
In this section we analyze this functional and prove that it
is, indeed, the large $B$ limit of MTF theory.

As before we assume that the confining potential $V$ is positive and that
$V(x) \rightarrow \infty$ as $\vert x \vert \rightarrow \infty$.
Moreover, we shall here assume that $V$ is continuous (in fact, we shall
make an even more stringent regularity assumption in Theorem 3.2 below).

We begin by showing the existence of a minimizer for (3.2). For general
continuous $V$ (without further assumptions) we must take into account the
possibility that the minimizing $\uprho$ may be a measure. In (3.2) we
therefore minimize over all positive measures $\uprho$ with $\int \uprho =
N$. It follows from the finite radius lemma given in the appendix that
$$E^{\rm C} (N,V) = \inf \left\{ \E^{\rm C} [\uprho ,V] : \ {\rm support} \
\uprho \subset \{ x: \vert x \vert \leq R_v \}, \ \uprho \geq 0, \ \int
\uprho = N \right\}. \eqno(3.3)$$ Here $R_v$ depends only on $v = V/N$.
Later on we shall show that the minimizer is, indeed, a function, and hence
that (3.2) does give us the large $B$ limit of MTF theory for suitable
$V$.

{\it 3.1. PROPOSITION (Existence and uniqueness of a minimizing measure).}
{\it Let $V$ be continuous. Then there is a unique positive measure
$\uprho^{\rm C}_{N,V}$ with $\int \uprho^{\rm C}_{N,V} = N$ such that
$E^{\rm C} (N,V) = \E^{\rm C} [\uprho^{\rm C}_{N,V} ; V]$.}

{\it Proof:} [Note: we write measures as $\uprho(x)\d x$, even if they are not
absolutely continuous with respect to Lebesgue measure.]
By (3.3) we can choose a sequence of positive measures,
$\uprho_1, \uprho_2, \dots$,
supported in
$\{ x: \vert x \vert \leq R_v \}$ with $\int \uprho_n = N$, such
that $\lim \nolimits_{n \rightarrow \infty} \E^{\rm C} [\uprho_n ; V] =
E^{\rm C}
(N,V)$.  The bounded measures are the dual of the continuous functions, and
so, by the Banach-Alaoglu Theorem, we may assume (by possibly passing to a
subsequence) that $\uprho_n$ converges weakly to a positive measure
$\uprho$ still supported in $\{ x: \vert x \vert \leq R_v \}$.  In particular
it follows that $\int \uprho = N$ and $\int \uprho_n V \rightarrow \int
\uprho V$.
Moreover, the product measure $\uprho_n \times \uprho_n \rightarrow \uprho
\times \uprho$ weakly.  Hence
$$\eqalignno{\int\int \uprho (x) \uprho (y) \vert x-y \vert^{-1} \d x \d y
&= \lim \limits_{\delta \rightarrow 0} \int\int \uprho (x) \uprho
(y) (\vert x-y \vert + \delta)^{-1} \d x \d y \cr
&= \lim \limits_{\delta \rightarrow 0} \lim \limits_{n \rightarrow \infty}
\int\int \uprho_n (x) \uprho_n (y) (\vert x-y \vert^{-1} +
\delta)^{-1} \d x \d y \cr
&\leq \liminf \limits_{n \rightarrow \infty} \int\int \uprho_n
(x) \uprho_n (y) \vert x-y \vert^{-1} \d x \d y. \qquad&(3.4) \cr}$$
The first equality follows by the Lebesgue's monotone convergence Theorem.
The last inequality is an immediate consequence of the pointwise bound
$(\vert x-y \vert + \delta)^{-1} \leq \vert x-y \vert^{-1}$.

We conclude from (3.4) that
$$E^{\rm C} (N,V) \leq \E^{\rm C} [\uprho ;V] \leq \liminf
\limits_{n \rightarrow \infty}
\E^{\rm C} [\uprho_n ; V] = E^{\rm C} (N,V) \eqno(3.5)$$
and hence that $\uprho$ is a minimizer.

The uniqueness of $\uprho$ follows from strict convexity of $D(\uprho,
\uprho)$. \hfill{\bf Q.E.D.}

The next theorem gives conditions, which are perfectly adequate for the
physical applications, under which the minimizer is a function and not just
a measure.  Moreover, that function has certain nice integrability
properties.

{\it 3.2 THEOREM (The minimizer is a function).} {\it Assume that the
potential $V$ is in the class $C^{1, \alpha}_{\rm loc}$ for some $0 <
\alpha \leq 1$.
(I.e., $V$ is once continuously differentiable and for each $R>0$ its
derivative satisfies
$$\vert \nabla V (x) - \nabla V(y) \vert \leq c_R \vert x-y \vert^\alpha,
\eqno(3.6)$$
inside the ball of radius $R$ centered at the origin
for some constant $c_R > 0$.) Then the minimizing measure
$\uprho^{\rm C}_{N,V}$ of Prop.~3.1 is a function. It has the properties
(with $\widehat{\uprho}$ being the Fourier transform of $\uprho$)
$$\eqalignno{\int \vert \widehat{\uprho}^{\rm C}_{N,V}(p) \vert^2 \vert p
\vert^r \d p < C_1, \qquad &-1 \leq r < \alpha \qquad&(3.7)\cr \uprho^{\rm
C}_{N,V}* \vert x \vert^{-1} \ {\rm is \ continuous} &\qquad&(3.8)\cr \int
\uprho^{\rm C}_{N,V} (x)^q \d x < C_2, \qquad &1 \leq q < {4 \over 2-
\alpha} \qquad&(3.9)\cr}$$ where $C_1$ and $C_2$ are constants (implicitly
computed below) that depend only on the constants $c_R, q, r, \alpha$ and on
$N$.}

{\it Proof:}  We write $\uprho^{\rm C}_{N,V}=\uprho$.
We know that $\int \uprho = N$ and that $\uprho$ has compact
support.  {F}rom the former fact we conclude that $\widehat{\uprho}$ is
well defined, continuous and bounded.

Let $g_a$ be the function with Fourier transform
$$\widehat g_a (p) = \int g_a (x) e^{i p x} \d x = \cases{1, &$\vert p
\vert \leq a$ \cr
0, &$\vert p \vert \geq a$\cr} \eqno(3.10)$$
Then $g_a$ is continuous, $\int g_a = 1$ and $\int y g_a (y) \d y = 0$.

Let $\uprho_a$ be the convolution $\uprho * g_a$, so that $\int \uprho_a = N$.

Since $\uprho$ is a minimizer,
$\E^{\rm C} [\uprho ; V]\leq \E^{\rm C} [\uprho_a ; V]$.  Explicitly this
inequality is
$$\int (\uprho_a - \uprho)V  + D (\uprho_a, \uprho_a) - D (\uprho, \uprho)
\geq 0, \eqno(3.11)$$
Since $\int V \uprho _a = \int (V * g_a) \uprho$ we can rewrite the first
term as
$$\int\int [V(x-y) - V(x)] \uprho (x) g_a (y) \d x \d y.$$
By integrating $\nabla V$ along the line from $x-y$ to $x$, and using
(3.6), we have
$$\vert V (x-y) - V(x) + y\cdot\nabla V (x) \vert \leq c \vert y \vert^{\alpha
+1}.$$
(Note that by the finite radius Lemma~A1 all integrals are restricted
to a finite ball.)
Using that $\int y g_a (y) = 0$ we can estimate the first term in (3.11)
as follows:
$$\int ( \uprho_a - \uprho)V \leq C \int \vert y \vert^{\alpha +1} \vert
g_a (y) \vert \d y = (\const.) a^{-\alpha -1}.$$

The last two terms in (3.11) are
$$(\const.) \int \vert \widehat{\uprho} (p) \vert^2 \vert p \vert^{-1}
[\vert \widehat g_a (p) \vert^2 - 1] \d p = (\const.) \int \limits_{\vert p
\vert \geq a} \vert \widehat{\uprho} (p) \vert^2 \vert p \vert^{-1} \d p$$
(Recall that $\widehat{\vert x \vert^{-1}} = (\const.) \vert p \vert^{-1}$
in two-dimensions.)  The inequality (3.11) thus implies
$$\int \limits_{\vert p \vert \geq a} \vert \widehat{\uprho} (p) \vert^2
\vert p \vert^{-1} \d p \leq (\const.) a^{-\alpha - 1}. \eqno(3.12)$$

Using (3.12) and $\widehat{\uprho} (p) \leq \int \uprho = N$ we can now
prove (3.7) as follows.
$$\eqalignno{\int \vert \widehat{\uprho} (p) \vert^2 \vert p \vert^r \d p
&= \int \limits_{\vert p \vert \leq 1} \vert \widehat{\uprho} (p) \vert^2
\vert p \vert^r \d p + \sum \limits^\infty_{n=0} \ \ \int \limits_{2^n \leq
\vert p \vert \leq 2^{n+1}} \vert \widehat{\uprho} (p) \vert^2 \vert p
\vert^r \d p \cr
&\leq N^2 \int \limits_{\vert p \vert \leq 1} \vert p \vert^r \d p +
(\const.) \sum \limits^\infty_{n=0} 2^{(r+1)(n+1)} \int \limits_{2^n \leq
\vert p \vert \leq 2^{n+1}} \vert \widehat{\uprho} (p) \vert^2 \vert p
\vert^{-1} \d p \cr
&\leq (\const.) N^2 + (\const.) \sum \limits^\infty_{n=0} 2^{(r+1)(n+1) -
n(\alpha +1)} < \infty, \cr}$$
if $r < \alpha$.

To prove that $\uprho * \vert x \vert^{-1}$ is continuous is now easy. We
simply prove that its Fourier transform is integrable.  The Fourier
transform of $\uprho * \vert x \vert^{-1}$ is $(\const.)
\widehat{\uprho} (p) \vert p \vert^{-1}$ and
$$\int \vert \widehat{\uprho} (p) \vert \vert p \vert^{-1} \d p \leq \int
\limits_{\vert p \vert \leq 1} \vert \widehat{\uprho} (p) \vert \vert p
\vert^{-1} \d p + \left( \int \limits_{\vert p \vert \geq 1}
\vert \widehat{\uprho} (p) \vert^2 \vert p
\vert^r \d p \right)^{1/2} \left( \int \limits_{\vert p \vert \geq 1} \vert
p \vert^{-r-2} \d p \right)^{1/2} < \infty.$$

Finally, we prove (3.9).  For $1 \leq q \leq 2$ there is no problem
because we know that $\int \uprho^2 = \int \widehat{\uprho^2}$. Hence
$\uprho$ is a square integrable function and, since $\int \uprho = N$, we
conclude by H\"older's inequality that (3.9) holds for $1 \leq q \leq 2$.
For $q > 2$ we will prove that
$$\int \vert \widehat{\uprho} \vert^t < \infty \ {\rm with}\ {4 \over
\alpha + 2} < t = {q
\over q-1} < 2.$$
This will prove (3.9) by the Hausdorff-Young inequality, which states
that $\left( \int \vert \widehat{\uprho} \vert^t \right)^{1/t} \geq \left(
\int \uprho^q \right)^{1/q}$ when $1 \leq t \leq 2$.

We write $\vert \widehat{\uprho} (p) \vert^t = \{ \vert \widehat{\uprho}
(p) \vert^t (1 + \vert p \vert)^m \} \{ (1 + \vert p \vert)^{-m} \}$ and
then use H\"older's inequality with $a^{-1} + b^{-1} = 1$ to conclude that
$$\int \vert \widehat{\uprho} \vert^t \leq \left[ \int \vert
\widehat{\uprho} (p) \vert^{ta} (1 + \vert p \vert)^{ma} \d p \right]^{1/a}
\left[ \int (1 + \vert p \vert)^{-mb} \d p \right]^{1/b}.$$
Thus $\int \vert \widehat{\uprho} \vert^t < \infty$ if we can satisfy $ta =
2, ma < \alpha$ and $mb > 2$, in addition to $a^{-1} + b^{-1} = 1$.  This
requires $\alpha /a > m > 2/b$, or $1 \leq a < 1 + \mfr1/2 \alpha$.  Thus,
we
require $t = 2/a > 4/(\alpha + 2)$ which, since $q = t/(t-1)$, means $q <
4/(2 - \alpha)$. \hfill {\bf Q.E.D.}

Corresponding to the minimization (3.2) there is a variational
``equation'' satisfied by the minimizer $\uprho$.
In the general case, in which $\uprho$ might be a measure, the variational
equation exists but is slightly complicated to state.

In physically
interesting cases $V$ is certainly in $C^{1, \alpha}$, in which case
Theorem 3.2 tells us that $\uprho$ is a function and that $\uprho * \vert x
\vert^{-1}$ is continuous.  Hence the total potential
$$V^{\rm C} = V + \uprho * \vert x \vert^{-1} \eqno(3.13)$$
is continuous.  It is then easy to derive by standard arguments, as in
Sect. 1, that $\uprho$ is the unique nonnegative solution to the
variational equation
$$\eqalignno{V(x) + \uprho * \vert x \vert^{-1} = \mu \qquad &{\rm if}
\quad \uprho (x) \not= 0 \cr
V(x) + \uprho * \vert x \vert^{-1} \geq \mu \qquad &{\rm if} \quad \uprho
(x) = 0, \qquad&(3.14)\cr}$$
for a unique $\mu > 0$.  As usual the chemical potential $\mu$ is a
monotone
function of the particle number $N = \int \uprho$.  In the special case of
a parabolic confining potential the solution to
(3.14) can be given in closed form.

{\it 3.3 PROPOSITION (Minimizer for the parabolic exterior potential).}
{\it
If $V(x) = K \vert x\vert^2$ then the minimizer of $\E^{\rm C}_{N,V}$ is
$$\uprho^{\rm C}_{N,V}(x) = \cases{{3 \over 2 \pi} N\lambda \sqrt{1 -
\lambda \vert x
\vert^2}, &if $\vert x \vert \leq \lambda^{-1}$ \cr
0, &if $\vert x \vert > \lambda^{-1}$ \cr} ,\eqno(3.15)$$
where $\lambda = (8K/3\pi N)^{2/3}$.
In fact, $\uprho^{\rm C}_{N,V}$
is the solution to (3.14) with $\mu = (3 \pi/4) N\lambda^{1/2}$.}

{\it Proof:\/}  This solution (3.15) was certainly known before; see e.g.,
[10].
We give the proof here for the convenience of the reader.
We only have to show that $\uprho=\uprho^{\rm C}_{N,V}$ is the solution to
(3.14).
It is enough to consider
the case $\lambda = 1$ and $N = 1$.  Then $V(x) = {3 \pi \over 8} \vert
x\vert^2$ and
$\mu = {3 \pi \over 4}$.  We may compute $\uprho * \vert x \vert^{-1}
= {3 \over 2 \pi} \int \sqrt{1 - \vert x-y \vert^2} \vert y \vert^{-1} \d
y$ by writing $y$ in polar coordinates $(\vert y \vert, \theta)$ and
performing the $\vert y \vert$ integration first.
$$\eqalignno{\uprho * \vert x \vert^{-1} &= {3 \over 2 \pi} \int
\int [(1 - \vert y \vert - \vert x \vert \cos \theta)^2 -
\vert x \vert^2 \sin^2 \theta]^{1/2} \d \theta \d \vert y \vert \cr
&= {3 \over 2 \pi} \int (1 - \vert x \vert^2 \sin^2 \theta)^{1/2}
\int \left( 1 - {(\vert y \vert - \vert x \vert \cos
\theta)^2 \over 1 - \vert x \vert^2 \sin^2 \theta} \right)^{1/2} \d \vert y
\vert \d \theta, \cr}$$
the integrations are over the intervals in $\theta$ and $\vert y \vert$ for
which the integrands are real.  Introducing the variable $t = (\vert y
\vert - \vert x \vert \cos \theta) (1 - \vert x \vert^2 \sin^2
\theta)^{-1/2}$ we obtain
$$\eqalignno{\uprho * \vert x \vert^{-1} &= {3 \over 2 \pi} \int
\limits^{\theta_m (x)}_{-\theta_m (x)} (1 - \vert x \vert^2 \sin^2
\theta) \d \theta \int^1_{-1} (1 - t^2)^{1/2} \d t, \cr
&= \mfr3/4 \int \limits^{\theta_m(x)}_{-\theta_m (x)} (1 - \vert x \vert^2
\sin^2 \theta) d \theta, \qquad&(3.16)\cr}$$
where
$$\theta_m (x) = \cases{\pi/2, &if $\vert x \vert \leq 1$ \cr
\sin^{-1} {1 \over \vert x \vert}, &if $\vert x \vert > 1$. \cr}$$
Thus
$$\uprho * \vert x \vert^{-1} = {3 \pi \over 4} - {3 \pi \over 8} \vert x
\vert^2 \quad {\rm if} \ \vert x \vert \leq 1$$
and
$$\uprho * \vert x \vert^{-1} \geq {3 \pi \over 4} - {3 \pi \over 8} \vert
x \vert^2 \quad {\rm if} \ \vert x \vert > 1$$
(The last inequality comes from the fact that the integral from $\theta_m$
to $\pi /2$ in (3.16) is negative.) \hfill {\bf Q.E.D.}

The energy function $E^{\rm C} (N,V)$ has the simple
{\bf scaling}:
$$E^{\rm C} (N,V) = N^2 E^{\rm C} \left( 1, {V \over N} \right).
\eqno(3.17)$$
The minimizing density $\uprho^{\rm C}_{N,V}$ for (3.2) scales as
$$\uprho^{\rm C}_{N,V}(x) = N \uprho^{\rm C}_{1,v} (x).$$
where $v = V/N$.

We shall now make precise in what sense the classical energy $E^{\rm C}$
is a limit of the MTF energy. In fact, in two different limits
(the large $B$ limit and the low coupling limit)  the MTF energy will
converge to the classical energy.
We first treat the large $B$ limit.

{\it 3.4 THEOREM (Large $B$ limit of MTF).}
{\it If the exterior potential $V$ is in the class $C^{1,\alpha}_{\rm loc}$
we have, as $B\to\infty$,
$$
	E^\MTF(N,B,V)\to E^{\rm C}(N,V)\eqno(3.18)
$$
and
$$
	\uprho^{\MTF}_{N,B,V}(x)\to \uprho^{\rm C}_{N,V},\eqno(3.19)
$$
in the weak $L^1$ sense.
}

{\it Proof:\/}
If we use $\uprho^{\MTF}_{N,B,V}$ as a trial density in $\E^{\rm C}$  and
recall that $j_B\geq 0$ we immediately obtain $E^{\rm C}(N,V)\leq
E^{\MTF}(N,B,V)$.

For the bound in the opposite direction we use
$\uprho^{\rm C}_{N,V}$ as a trial density for $\E^{\MTF}$.
In order to do this it is, however, important that we know (from
Theorem~3.2)
that $\uprho^{\rm C}_{N,V}$ is a function. Hence $j_B(\uprho^{\rm
C}_{N,V})$
is well defined. Moreover, from the definition of $j_B$,
$j_B(\uprho^{\rm C}_{N,V})\to0$ almost everywhere as $B\to\infty$
and $j_B(\uprho^{\rm C}_{N,V})\leq j_0(\uprho^{\rm C}_{N,V})$.
Since $j_0({\uprho})=(\pi/2)({\uprho})^2$
we know from (3.9) that $j_0(\uprho^{\rm C}_{N,V})$ is
integrable. The limit in (3.18) is therefore an immediate consequence
of Lebesgue's dominated convergence theorem.

The convergence of the densities in (3.19) follows in a standard way by
replacing
$V$ by $V+\varepsilon f$ with $f$ a bounded (measurable) function
and differentiating w.r.t. $\varepsilon$, see e.g., [29]. \hfill {\bf Q.E.D.}

We point out that if $\uprho^{\rm C}_{N,V}$ is a bounded function, as  it
is, e.g.,
for $V=K\vert x\vert^2$, then
$$
	E^\MTF(N,B,V)=E^{\rm C}(N,V),\eqno(3.20)
$$
for $B$ large enough,
because in that case $j_B(\uprho^{\rm C}_{N,V})$ vanishes for $B$ large.

Finally,
we now discuss the weak coupling limit in the case of homogeneous
exterior potentials.
Suppose $V$ is a homogeneous function of $x$,
$V(\lambda x)=\lambda^s V(x)$, $s>0$. If we consider the exterior
potentials
$kV(x)$ with $k>0$ the classical energy and density obey the scalings
$$
	E^{\rm C}(N,kV)= k^{1/(s+1)}E^{\rm C}(N,V)\eqno(3.21)
$$
and
$$
	\rho^{\rm C}_{N,kV}(x)=k^{2/(s+1)}
	\rho^{\rm C}_{N,V}\left(k^{1/(s+1)}x\right).\eqno(3.22)
$$

If $k$ is small we
see from (3.22) that the minimizing density for the MTF functional
will spread out and its kinetic energy will be negligible compared with the
classical terms. We prove this rigorously now.

{\it 3.5 THEOREM (Weak coupling limit of MTF with homogeneous potentials).}
{\it Let $V$ be $C^{1,\alpha}_{\rm loc}$ and homogeneous of degree $s$.
If $k\to0$
then
$$
	{E^\MTF(N,B,kV)\over E^{\rm C}(N,kV)}\to1,\eqno(3.23)
$$
and
$$
	k^{-2/(s+1)}\rho^\MTF_{N,B,kV}
	\left(k^{-1/(s+1)}x\right)
	\to\rho^{\rm C}_{N,V}\left(x\right),\eqno(3.24)
$$
in weak $L^1$ sense.
Both limits are uniform in $B$.
}

{\it Proof:}
As above we may use $\uprho^{\MTF}_{N,B,V}$ as a trial density in $\E^{\rm
C}$  to
conclude that $E^{\rm C}(N,V)\leq E^{\MTF}(N,B,V)$.

To prove the bound in the opposite direction we again use $\uprho^{\rm
C}_{N,V}$
as a trial density for $\E^{\MTF}$.
We then obtain from (2.15) and the scaling (3.22) that
$$
	E^{\MTF}(N,B,kV)\leq k^{2/(s+1)}\int j_0({\uprho}^{\rm C}_{N,V})
	+ E^{\rm C}(N,kV), \eqno(3.25)
$$
where we used that $j_b\leq j_0$.
If we compare this with the scaling in (3.21) we see that
$E^{\MTF}/E^{\rm C}\to1$
as $k\to0$ since $j_0({\uprho}^{\rm C}_{N,V})=(\pi/2)({\uprho}^{\rm
C}_{N,V})^2$ is integrable.

The convergence of the densities follows again by replacing
$V$ by $V+\varepsilon f$ and differentiating w.r.t. $\varepsilon$. \hfill{\bf
Q.E.D.}

In the same way as for the large $B$ limit (3.23) becomes an identity for small
$k$ if
$\rho^{\rm C}_{N,V}$ is a bounded function.

We may of course also introduce the scaling $V=Nv$ when $v$ is homogeneous
of degree $s$. Then $kV=Kv$, where $K=kN$, and the limit in (3.23) is
uniform
in $N$. The limit in (3.24) is uniform if we formulate it as $$
N^{-1}k^{-2/(s+1)}\rho^\MTF_{N,B,kV} \left(k^{-1/(s+1)}x\right)
\to\rho^{\rm C}_{1,v}\left(x\right).\eqno(3.26) $$

We remark that if a potential $W$ is asymptotically homogeneous in
the sense that there is a homogeneous
potential $V$ with
$\lim_{|x|\to\infty}W(x)/V(x)=1$, then
$$\lim_{k\to 0}E(N,B,kW)/k^{1/(1+s)}=E^{\rm C}(N,V)\eqno(3.27)$$
uniformly in $B$, where $s$ is the degree of homogeneity of $V$.

\bigskip
\noindent
{\bf IV.  THE CLASSICAL POINT CHARGE MODEL:  A LIMIT OF QUANTUM MECHANICS}

Another model that sheds some light on the physics of our problem --- and
that will also be important for bounding the difference between the TF
theory and the original quantum theory in Sect.~V --- is the classical
particle
model.  In this model the kinetic energy is simply omitted altogether, but
the point-like nature of the electrons is retained.

{\it 4.1.  DEFINITION (Classical particle energy).}  With $V(x)$ being
the confining potential the {\bf
classical particle energy} for $N$ points in $\R^2$ is defined by
$$\E^{\rm P} (x_1, \dots, x_N;V) =  \sum \limits^N_{i=1} V(x_i) + \sum
\limits_{1 \leq i < j \leq N} \vert x_i - x_j \vert^{-1}. \eqno(4.1)$$
The {\bf minimum classical particle energy} for $N$ point particles in
$\R^2$ is
$$E^{\rm P} (N,V) = \inf \{ \E^{\rm P} (x_1, \dots , x_N;V) : x_i \in \R^2 \}.
\eqno(4.2)$$

We shall estimate the particle energy $E^{\rm P} (N,V)$ in terms of the
classical
continuum energy $E^{\rm C} (N,V)$.  We first show that $E^{\rm C} (N,V)$ gives
an exact upper
bound on $E^{\rm P} (N,V)$.

{\it 4.2.  LEMMA (Upper bound for $E^{\rm P}$).}  {\it For all $N$ we have
$$E^{\rm P} (N,V) \leq E^{\rm C} (N,V) -  N^{3/2}/(8R_v), \eqno(4.3)$$
where $R_v$  is the maximal radius given in Lemma A1.}

{\it Proof:}  First, let us give a very simple argument that yields an
error term proportional to $N$ instead of $N^{3/2}$.
The energy $E^{\rm P} (N,V)$ is bounded above by
$$\int \E^{\rm P} (x_1,
\dots , x_N;V) \Phi (x_1, \dots , x_N)\d x_1 \dots \d x_N,$$ for any
nonnegative
function $\Phi$ with $\int \Phi=1$.
We take $\Phi (x_1, \dots , x_N) = \prod \limits^N_{i=1} {\uprho_{[1]}
(x_i)}$, where for simplicity we have introduced the notation
$\uprho_{[1]}$ for the minimizer
$\rho^{\rm C}_{1,V/N}$ for $\E^{\rm C} [\uprho ;V/N]$ with
$\int \rho^{\rm C}_{1,V/N} =1$. Note that $\uprho_{[1]}$ depends only on
$v=V/N$.
We  obtain
$$\eqalignno{E^{\rm P} (N,V) &\leq \int \E^{\rm P} (x_1, \dots , x_N;V) \prod
\limits^N_{i=1} {\uprho_{[1]} (x_i) } dx_1 \dots dx_N \cr
&= N \int V(x) \uprho_{[1]} (x) \d x + {N (N-1) \over 2} \int\!\!\int
\uprho_{[1]} (x)
\vert x-y \vert^{-1} \uprho_{[1]} (y) \d x \d y. \cr} $$
Recalling that the minimizer of $\E^{\rm C}$ is
$\uprho^{\rm C}_{N,V}(x)=N\uprho_{[1]} (x)$, we get
an error term $-aN$, with
$a = \mfr1/2 \int \int \uprho_{[1]} (x) \uprho_{[1]} (y)$ $\vert x-y
\vert^{-1} \d x \d y$.

Now we turn to a proof of (4.3) which, obviously, has to be more
complicated than the previous discussion.  By Lemma A1 there is a
fixed square, $Q$, centered at the origin, whose width, $W$, equals $2
R_v$,
such that the minimizer $\uprho=\uprho^{\rm C}_{N,V}$ for $\E^{\rm C}$ is
supported in $Q$.
For simplicity we suppose that $\sqrt{N}$
is an integer; if this is not so the following proof can be modified in an
obvious way.

First, cut $Q$ into $\sqrt{N}$ vertical, disjoint strips, $S_1, S_2, \dots
, S_{\sqrt{N}}$ such that $\int_{S_j} \uprho = \sqrt{N}$ for all $j$.  Let
$t_j$ denote the width of $S_j$, so that $\sum \nolimits^{\sqrt{N}}_{j=1}
t_j = W$.  Next, make $\sqrt{N} -1$ horizontal cuts in each $S_j$ so that the
resulting rectangles $R_{jk}$ for $k=1, \dots , \sqrt{N}$ satisfy
$\int_{R_{jk}} \uprho = 1$.  Denote the height of these rectangles by
$h_{jk}$, so that $\sum \nolimits^{\sqrt{N}}_{k=1} h_{jk} = W$ for each
$j$.  Having done this we note, by convexity, that for each $j$
$$N^{-1/2} \sum \limits^{\sqrt{N}}_{k=1} (t_j + h_{jk})^{-1} \geq \left[
N^{-1/2} \sum \limits^{\sqrt{N}}_{k=1} (t_j + h_{jk}) \right]^{-1} =
\bigl[ t_j + N^{-1/2} W\bigr]^{-1}.$$
Again, using the same convexity argument for the $j$-summation, we have
that
$$\sum \limits^{\sqrt{N}}_{j=1} \sum \limits^{\sqrt{N}}_{k=1} (t_j +
h_{jk})^{-1} \geq {N^{3/2} \over 2W}. \eqno(4.4)$$

Let $\uprho_{jk}$ be the minimizing density $\uprho$ restricted to the
rectangle $R_{jk}$, i.e. $\uprho_{jk} (x) = 1$ if $x \in R_{jk}$ and $=0$
otherwise.  Thus, $\int \uprho_{jk} = 1$.  We denote these $N$ function by
$\uprho^i, i = 1, \dots , N$.  Define $\Phi (x_1, \dots, x_N) := \prod
\limits_{i=1}^N \uprho^i (x_i)$ and, as in the previous proof, a simple
computation yields
$$E^{\rm P} (N,V) \leq \int \E^{\rm P} \Phi = \E^{\rm C} [\uprho ;V] - \sum
\limits^N_{i=1}
D (\uprho^i, \uprho^i),$$
with $D (f,g) = \mfr1/2 \int\int f(x) g(y) \vert x-y \vert^{-1} \d x \d y$.

To complete our proof we note that as long as $x$ and $y$ are in $R_{jk}$
we have that $\vert x-y \vert^{-1} \geq (t_j + h_{jk})^{-1}$.  Thus, $\sum
\limits_{j,k} D (\uprho_{jk}, \uprho_{jk}) \geq N^{3/2}/4W$ by (4.4) and
the fact that $\int \uprho_{jk} = 1$. \hfill {\bf Q.E.D.}

{\it 4.3.  LEMMA (Lower bound for $E^{\rm P}$).}  {\it Assume that
$V$ is a potential in $C^{1,\alpha}_{\rm loc}$. Then for all $N$ we have
$$
	E^{\rm P} (N,V)\geq E^{\rm C} (N,V)-bN^{3/2}.\eqno(4.5)
$$
with
$$
	b=\mfr4/3\sqrt{\mfr2/3}\int\left(\uprho^{\rm C}_{1,v} (x)\right)^{3/2}\d
x+
	\left({2\pi\over 2-p}\right)^{1/p}(2R_v)^{-1+(2/p)}
	\left( \int \left(\uprho^{\rm C}_{1,v}\right)^q
\right)^{1/q}
	\eqno(4.6)
$$
and where $q$ is any number satisfying $2 < q < 4/(2-\alpha)$, $R_v$ is the
maximal radius given  in Lemma A1 and $p=q/(q-1)<2$.
As we explained in Theorem 3.2, our hypothesis that $V\in C^{1,\alpha}_{\rm
loc}$
implies that for $q < 4/(2-\alpha), \left( \int
\left(\uprho^{\rm C}_{1,v}\right)^q
\right)^{1/q}$ is less than some constant that depends only on $q,\alpha$
and $v = V/N$.  In particular, $b$ depends only on $v$.
}

We see from (4.3) and (4.5) that when $V \in C^{1,\alpha}_{\rm loc}$ the power
3/2
in the error term is optimal.

In order to prove (4.5) we need the following
electrostatics lemma of Lieb and
Yau [26]. The original version was for ${\bf R}^3$; we state it here
for ${\bf R}^2$ solely for the convenience of our present application.

{\it 4.4. LEMMA (The interaction of points and densities).}  {\it
Given points $x_1, \ldots, x_N$ in ${\bf R}^2$, we define Voronoi
cells $\Gamma_1,\ldots, \Gamma_N\subset {\bf R}^2$ by
$$
	\Gamma_j=\{y\in {\bf R}^2 : |y-x_j|\leq |y-x_k|\ \
	\hbox{for all \ } k\ne j\} \quad.
$$
These $\Gamma_j$ have disjoint interiors  and their union covers
${\bf R}^2$. We also define $R_j$ to be the distance from $x_j$
to the boundary of $\Gamma_j$, i.e., $R_j$ is half the distance
of $x_j$ to its nearest  neighbor. Let $\uprho$ be any (not necessarily
positive) function on ${\bf R}^2$. (In general, $\uprho$ can be replaced
by a measure, but it is not necessary for us to do so.) Then
(with $D(f,g)=\mfr1/2\int\!\!\int f(x)g(y)|x-y|^{-1}\d x\d y$ )
$$
	\eqalignno{\sum_{1\leq i<j\leq N} |x_i-x_j|^{-1}\geq&
	- D(\uprho, \uprho)
	+ \sum_{j=1}^N\int_{{\bf R}^2} \uprho(y)|y-x_j|^{-1}\d y\cr
	&{}+\mfr1/8 \sum_{j=1}^N R_j^{-1}
	-\sum_{j=1}^N \int_{\Gamma_j}\uprho(y) |y-x_j|^{-1}\d y\quad.
	&(4.7)}
$$
}

{\it Proof of (4.5):}  We choose $\uprho$ in (4.7) to be the minimizer
$\uprho^{\rm C}_{N,V}$
for the functional $\E^{\rm C}$ and we choose the $x_i$'s to be any
(not necessarily minimizing) configuration for $\E^{\rm P}$. It is
important, for us that a minimizer exists for $\E^{\rm C}$, for then
$\uprho$ satisfies the Euler-Lagrange equation (3.14).
Since $\int\uprho=N$, we conclude from  (3.14) that
$$
	\sum_{j=1}^N\left[\int\uprho(y)|y-x_j|^{-1}\d y+V(x_j)\right]
	=\sum_{j=1}^N V^{\rm C}(x_j)\geq N\mu=\int V^{\rm C}\uprho
	=2 D(\uprho, \uprho) +\int V\uprho\quad.
$$
Thus, if we add $\sum_j V(x_j)$ to both sides of (4.7)
we have that
$$
	E^{\rm P} (N,V)\geq E^{\rm C} (N,V)+ \sum_{j=1}^N\mfr1/8 R_j^{-1}
	-\int_{\Gamma_j}\uprho(y)|y-x_j|^{-1}\d y\quad.\eqno(4.8)
$$
Our goal will be to control the rightmost term in (4.8)
by the $R_j^{-1}$ term.

We split each region $\Gamma_j$ into two disjoint subregions,
$\Gamma_j=A_j\cup B_j$, where
$$
	A_j:=\{x : |x-x_j|< R_j\},\quad
	B_j:=\{x\in \Gamma_j : |x-x_j|\geq R_j\}\quad.
$$
Then, by H\"older's inequality
$$
	\eqalignno{\sum_{j=1}^N\int_{B_j} |y-x_j|^{-1}\uprho(y)\d y&\leq
	\left(\sum_j\int_{\Gamma_j}\uprho^{3/2}\right)^{2/3}\left(
	\sum_j\int_{|y-x_j|\geq R_j}|y-x_j|^{-3}\d y\right)^{1/3}\cr
	&=\left( \int \uprho^{3/2} \right)^{2/3}
\left(2\pi\sum_j R_j^{-1}\right)^{1/3}\quad.
	&(4.9)}
$$

If we define $X:=\sum_j R_j^{-1}$ we can rewrite (4.9) plus the
$X/8$ term in (4.8) as $\mfr1/8 X- \left( \int \uprho^{3/2} \right)^{2/3}
X^{1/3}$.
The minimum of this quantity, over all values $X$, is
$(4/3)\sqrt{2/3} \int \uprho^{3/2}$,
and thus we have accounted for the first error term in (4.6).

To estimate the term $I:=\sum_{j=1}^N\int_{A_j}\uprho(y)|y-x_j|^{-1}\d y$
some control is needed over the possible singularities of $\uprho$.
Let $p=q/(q-1)$ be the dual of $q$. Then
$$
	I\leq \left( \int \uprho^q \right)^{1/q}
\left[\sum_j\int_{A_j}|y-x_j|^{-p}\d y\right]^{1/p}
	= \left( \int \uprho^q \right)^{1/q}
\left[{2\pi\over 2-p}\sum_j R_j^{2-p}\right]^{1/p}\quad.
	\eqno(4.10)
$$

We note that, since $1\leq p< 2$,
$$
	\left[\sum_{j=1}^N R_j^{2-p}\right]^{1/p}\leq
	\left[ \sum_{j=1}^N R_j^{2}\right]^{1/p-1/2}N^{1/2}
$$
by H\"older's inequality. Now $\pi R_j^2$ is the area of the disc
$A_j$ and thus $\pi\sum_j R_j^2$ is the total area of all these
disjoint discs.  How large can this area be? To answer this we recall
Lemma A1 in the appendix which states that for the purpose
of finding a set of points that minimizes the classical
particle energy $\E^{\rm P}$
we can restrict attention to
a disc of radius $R_v$, centered at the origin.
We may therefore assume that our $x_j$'s satisfy $|x_j|\leq R_v$.
This we can do whether or not an energy minimizing configuration
exists. Having done so and assuming that $N\geq 2$ we have that
$R_j<R_v$ for all $j$, and hence all
our discs are contained in a disc of
radius $2R_v$ centered at the origin . Thus $\sum_j R_j^2\leq
(2R_v)^2$,
and our second error term, (4.10), is bounded above by
$\left( \int \uprho^q \right)^{1/q} ({2\pi\over 2-p})^{1/p}(2R_v)^{-1+2/p}
N^{1/2}$.
This yields (4.6). \hfill{\bf Q.E.D.}

\bigskip
\noindent
{\bf V.  MTF THEORY IS THE HIGH DENSITY LIMIT OF
QUANTUM MECHANICS}

In this section we prove that the quantum energy and the quantum density
are given by the corresponding MTF quantities to leading order for large
$N$.
These are the statements of Theorems~1.1 and~1.2.
We shall not prove Theorem~1.2 since it follows from Theorem~1.1
in a standard way by replacing
$V$ by $V+\varepsilon f$ with $f$ a function in $C^{1,\alpha}_{\rm loc}$
and differentiating w.r.t. $\varepsilon$, see e.g., [29].

We shall prove Theorem~1.1 by giving sharp upper and lower bounds to the
quantum
ground state energy.  The upper bound is obtained by a variational
calculation using the magnetic coherent states introduced in [25,21].  The
lower bound is more difficult.  Besides the results of the previous
sections several ingredients are needed.  The first ingredient is a kinetic
energy inequality of the Lieb-Thirring type [31-32], [21], [33].  Such an
inequality estimates the kinetic energy of a many-body wave function from
below in terms of a functional of the density.  The proof of this
inequality in the 2-dimensional case considered here is harder than the
3-dimensional case treated in [21] and involves some new mathematical
ideas.  The second ingredient is a lower bound on the exchange-correlation
energy.  The proof of this inequality is similar to that given in [34-35]
for the 3-dimensional case.

Once the kinetic energy and exchange-correlation inequalities have been
established the proof of the lower bound is completed by a coherent states
analysis.

We start by discussing the magnetic coherent states used in the proofs of
both the upper and lower bounds.  They are constructed from the kernels
$$\Pi_{\alpha \sigma} (x \sigma^\prime, y \sigma^{\prime\prime}) = {B
\over 2 \pi} \exp \{ i(x \times y) \cdot B - \vert x-y \vert^2 B/4\}
L_\alpha
(\vert x-y \vert^2 B/2) \delta_{\sigma \sigma^\prime} \delta_{\sigma
\sigma^{\prime\prime}} \eqno(5.1)$$
of the projection operators onto the Landau levels $\alpha = 0,1,2, \dots$
with
$z$-component of spin $\sigma = \pm 1/2$.  Here $L_\alpha$ are Laguerre
polynomials normalized by $L_\alpha (0) = 1$.  In fact, all that matters
are the projectors $\Pi_\nu$ on the states with energy $\varepsilon_\nu
(B)$; these are given by a sum of at most two of the projections
$\Pi_{\alpha
\sigma}$.  More precisely,
$$\Pi_\nu = \sum \limits_{\scriptstyle \alpha, \sigma \atop \scriptstyle
\alpha + \mfr1/2 + \gamma \sigma = \varepsilon_\nu (B)/B} \Pi_{\alpha
\sigma}
.\eqno(5.2)$$

We shall not need the explicit form (5.1).  The three important properties
of $\Pi_\nu$ that we use are the following
$$\eqalignno{\sum \limits_\nu \Pi_\nu (x \sigma^\prime,
y \sigma^{\prime\prime}) &= \delta (x-y) \delta_{\sigma^\prime
\sigma^{\prime\prime}}, \qquad&(5.3)\cr
\sum \limits_{\sigma^\prime} \Pi_\nu (x\sigma^\prime, x
\sigma^\prime) &= d_\nu (B) \qquad&(5.4)\cr
H_{\rm kin} \Pi_\nu &= \varepsilon_\nu (B) \Pi_\nu, \qquad&(5.5)\cr}$$
where $H_{\rm kin}$ is given by (2.1).

Let $g$ be a real continuous function on $\R^2$, with $g(x) = 0$ for $\vert
x
\vert > 1, \int g^2 = 1$ and $\int (\nabla g)^2 < \infty$.  (The optimal
choice that minimizes $\int (\nabla g)^2$ is the Bessel function $J_0$,
suitably scaled and normalized.)  Define $g_r (x) = r^{-1} g(x/r)$, with $0
< r < 1$ to be specified later.  For each $u \in \R^2, \nu = 0,1,2$, we
define
the operator $\Pi_{\nu u}$ --- the coherent ``operator'' --- with kernel
$$\Pi_{\nu u} (x \sigma^\prime, y \sigma^{\prime\prime}) = g_r (x-u)
\Pi_\nu (x \sigma^\prime, y \sigma^{\prime\prime}) g_r (y-u). \eqno(5.6)$$
It easily follows from (5.3,4) and the properties of $g$ that these kernels
satisfy the coherent operator identities ([36])
$$\sum \nolimits_\nu \int \Pi_{\nu u}(x \sigma^\prime, y
\sigma^{\prime\prime}) \d u = \delta (x-y) \delta_{\sigma^\prime
\sigma^{\prime\prime}} \eqno(5.7)$$
$$\Tr \,\Pi_{\nu u}= \sum \limits_\sigma \int \Pi_{\nu u} (x \sigma^\prime,
x
\sigma^\prime) \d x = d_\nu (B). \eqno(5.8)$$
Moreover, a simple computation gives, using (5.5),
$$\eqalignno{\Tr [H_{\rm kin} \Pi_{\nu u}] &= d_\nu (B) [\varepsilon_\nu
(B) +
\int (\nabla g_r)^2], \qquad&(5.9) \cr
\Tr [V \Pi_{\nu u}] &= d_\nu (B) V * g_r^2 (u), \qquad&(5.10)\cr}$$
where $V$ is a (continuous) potential and $*$ denotes convolution.
Likewise, for all $f$ with $\langle f \vert f \rangle = 1$
$$\eqalignno{\langle f \vert H_{\rm kin} \vert f \rangle &= \sum
\limits_\nu \int
\varepsilon_\nu (B) \langle f \vert \Pi_{\nu u} \vert f \rangle \d u - \int
(\nabla g_r)^2 \qquad&(5.11)\cr
\langle f \vert V * g^2_r \vert f \rangle &= \sum \limits_\nu \int V(u)
\langle f \vert \Pi_{\nu u} \vert f \rangle \d u. \qquad&(5.12)\cr}$$
Equations (5.9,10) will be used in proving the upper bound, while (5.11,12)
are needed for the lower bound.
\bigskip\noindent
{\bf 5.1 THE UPPER BOUND}

We use the variational principle of [37].  According to this
principle
$$E^{\rm Q} (N,B,V) \leq \Tr [(H_{\rm kin} + V)K] + \mfr1/2 \sum
\limits_{\sigma^\prime,
\sigma^\prime} \int_{\R^2} \int_{\R^2} {K(x\sigma, x \sigma)
K(y\sigma^\prime,
y \sigma^\prime) \over \vert x-y \vert} \d x \d y,\eqno(5.13)$$
for all operators $K$, with kernel $K (x \sigma, y \sigma^\prime)$,
satisfying
$$0 \leq \langle f \vert K \vert f \rangle \leq \langle f \vert f
\rangle, \eqno(5.14)$$
for all $f$, and
$$\Tr [K] = \sum \nolimits_\sigma \int_{\R^2} K(x \sigma, x \sigma) \d x =
N.
\eqno(5.15)$$
We shall choose $K$ as follows.  Let $\uprho^{\MTF}$ be the MTF
density, i.e., the minimizer of the functional (1.5) with $\int
\uprho^{\MTF} = N$.  Denote by $\nu_{\max} (x)$ the highest filled level.
Then
$$0 \leq \uprho^{\MTF} (x) - \sum \limits_{\nu \leq \nu_{\max} (x)} d_\nu
(B) < d_{\nu_{\max} (x) + 1} (B). \eqno(5.16)$$
We introduce the filling factors
$$f_\nu (x) = \cases{1 &$\nu \leq \nu_{\max} (x)$ \cr
[\uprho^{\MTF} (x) - \sum \limits_{\nu \leq \nu_{\max} (x)} d_\nu
(B)]/d_{\nu_{\max} (x) + 1} (B), &$\nu = \nu_{\max} (x) + 1$ \cr
0 &$\nu > \nu_{\max} (x) + 1$ \cr}. \eqno(5.17)$$
and define
$$K (x \sigma, y \sigma^\prime) = \sum \limits_\nu
\int f_\nu (u) \Pi_{\nu u} (x
\sigma, y \sigma^\prime)du, \eqno(5.18)$$
with $\Pi_{\nu u}$ as in (5.6).  It follows from (5.7) that $K$ satisfies
(5.14) and from (5.8), (5.16) and (5.17) that $\Tr [K] = \int \uprho^{\MTF} (u)
\d
u
= N$.

Note that (5.4) and (5.6) imply $$\sum \limits_\sigma K(x \sigma, x \sigma)
= \uprho^{\MTF} * g^2_r (x). \eqno(5.19)$$ Hence, the last term in (5.13)
is $D(\uprho^{\MTF} * g^2_r, \uprho^{\MTF} * g^2_r)$, where the functional
$D$ was defined in (1.4). By convexity of $D$ we find that
$$D(\uprho^{\MTF} * g^2_r, \uprho^{\MTF} * g^2_r) \leq D(\uprho^{\MTF},
\uprho^{\MTF}).$$ {}From (5.9), (5.10) and (5.13) we obtain
$$\eqalignno{E^{\rm Q} (N,B,V) &\leq \E^{\MTF} (\uprho^{\MTF}) + N \int
(\nabla g_r)^2 \d x + \int [V * g^2_r (*) - V(x)] \uprho^{\MTF} (x) \d x
\cr &\leq E^{\MTF} (N,V,B) + N r^{-2} \int (\nabla g(x))^2 \d x + N^2 \sup
\limits_{\vert x \vert < R} [v * g^2_r (x) - v(x)], \qquad&(5.20)\cr}$$
where $R = R_v$ is the finite radius and we have written $V = Nv$. Since
$v$ is in $C^{1, \alpha}$ $$\sup \limits_{\vert x \vert < R} \vert v *
g^2_r (x) - v(x) \vert \leq (\const.) r.$$ We can choose $r = r_N$ such
that $r_N \rightarrow 0$ and $r^{-2}_N/N \rightarrow 0$ as $N \rightarrow
\infty$. This means that $r_N$ should go to zero but still be large
compared with the average spacing $N^{-1/2}$ between electrons. The optimal
choice is of the order $r=(\const.)N^{-1/3}$. Thus the error
$$[E^{\rm Q} (N,B,V) - E^{\MTF} (N,B,V)] N^{-2}$$ is bounded above by a
function
$\varepsilon^+_N (v)=c^+(v)N^{-1/3}$ (independent of $B$). This finishes
the proof of the upper bound.

{\it 5.2 THEOREM (Lieb-Thirring inequality in 2 dimensions).}

{\it  Let
$H_{\A}={1\over 2}(i\nabla-{\bf A})^2+{\bf S}\cdot {\bf B}$. (This is the
operator $H_{\rm kin}$ from (2.1) with $\gamma=1$.) Let $W$ be a
locally integrable
function and denote by
$e_1(W), e_2(W), \ldots$ the negative eigenvalues (if any)
of the operator $H=H_{\bf A}-W$ defined on
$L^2({\bf R}^2;{\bf C}^2)$, the space of wave functions of a single
spin-$\mfr1/2$ particle.  Define $\vert W \vert_+ (x) = \mfr1/2 [\vert W(x)
\vert + W(x)]$.  For all $0<\lambda<1$
we then have the estimate
$$
	\sum_j |e_j(W)|\leq  \lambda^{-1}{B \over 2\pi} \int_{\R^2} |W|_+
(x) \d x + \mfr3/4(1-\lambda)^{-2} \int_{\R^2} |W|_+^2 (x) \d x.
$$
}

{\it Proof:} For any self-adjoint operator $A$ we denote by
$N_{\alpha}(A)$ the number of eigenvalues of $A$ greater than or equal
to $\alpha$.

Since replacing $W$ by its positive part $|W|_+$ will only
enhance the sum of the negative eigenvalues we shall henceforth assume that
$W$ is positive, i.e., $W=|W|_+$. We consider the Birman-Schwinger kernel
$$
	    K_E=W^{1/2} ( H_{\bf A}+ E)^{-1}
		W^{1/2} .
$$
According to the Birman-Schwinger principle (see, e.g., [38], p.89)
the number $N_E(-H)$ of eigenvalues
of $H$ below $-E$ is equal to the number $N_1(K_E)$ of eigenvalues of $K_E$
greater than or equal to 1. We find
$$
	\sum_j |e_j(W)| = \int_0^{\infty} N_E(-H) \d E =\int_0^{\infty}
	N_1(K_E) \d E .
$$

In order to estimate $N_1(K_E)$ we decompose the Birman-Schwinger kernel
into a part $K_E^0$ coming from the lowest Landau level and
a part $K_E^>$ coming from the higher levels. If
$\Pi_0$ is the projection onto the lowest Landau band these two parts are
defined by
$$
K_E^0=W^{1/2}
	\Pi_0 ( H_{\bf A}+ E)^{-1}\Pi_0
		W^{1/2}
	=E^{-1}W^{1/2}\Pi_0 W^{1/2}\eqno(5.21)
$$
and
$$
	K_E^>=W^{1/2}(I-\Pi_0) ( H_{\bf A}+ E)^{-1}(I-\Pi_0)
						W^{1/2}.
$$
Since $\Pi_0$ commutes with $H_{\bf A}$ we have $K_E=K_E^0+K_E^>$.

Now we use Fan's theorem [39],
which states that if $\mu_1(X)\geq\mu_2(X)\geq\ldots$ denote the
eigenvectors of a self-adjoint compact operator $X$ then
$\mu_{n+m+1}(X+Y)\leq \mu_{n+1}(X)+\mu_{m+1}(Y)$ for $n,m\geq0$.
{}From this we have $N_1(X+Y)\leq N_{\lambda}(X)+N_{1-\lambda}(Y)$
which, in our case, reads as follows.
$$
	N_1(K_E)\leq N_{\lambda}(K_E^0)+N_{1-\lambda}(K_E^>), \ \hbox{for}\
0\leq\lambda\leq1.
$$
This inequality permits us to
consider the two parts of $K_E$ separately. We first consider the
contribution from the lowest level:
$N_{\lambda}(K_E^0)=N_{\lambda E}(W^{1/2}\Pi_0 W^{1/2})$.
We get
$$
	\eqalign{\int_0^{\infty} N_{\lambda}(K_E^0) \d E
	= &\int_0^{\infty}
   N_{\lambda E}(W^{1/2}\Pi_0 W^{1/2}) \d E \cr =&
	\lambda^{-1} \int_0^{\infty} N_E (W^{1/2}
	\Pi_0 W^{1/2}) \d E = \lambda^{-1} {\rm Tr}
	(W^{1/2}\Pi_0 W^{1/2})\cr = & \lambda^{-1}
	\int \Pi_0(x,x) W(x) \d x =
	\lambda^{-1} {B \over 2 \pi} \int W(x) \d x . }
$$
The second part is straightforward. We first notice that
$H_{\bf A}(I-\Pi_0)\geq B(I-\Pi_0)$. Hence
$H_{\bf A}(I-\Pi_0)\geq \mfr2/3(H_{\bf A}+\mfr1/2B)(I-\Pi_0)\geq
\mfr1/3(i\nabla-{\bf A})^2(I-\Pi_0)$. (Note that $(i\nabla-{\bf A})^2$
commutes with
$\Pi_0$.) Since the operator inequality $0<X\leq Y$ implies
$X^{-1}\leq Y^{-1}$ we have that
$$
	K_E^{>}\leq W^{1/2}(I-\Pi_0)[\mfr1/3(i\nabla-
        {\bf A})^2+E]^{-1} (I-\Pi_0)W^{1/2}
	\leq W^{1/2}[\mfr1/3(i\nabla-
        {\bf A})^2+E]^{-1} W^{1/2}.
$$
We conclude that  $N_{1-\lambda}(K_E^>)
\leq N_1(\widetilde{K}_E)$, where
$$
	\widetilde{K}_E=\left((1-\lambda)^{-1}W\right)^{1/2}[\mfr1/3(i\nabla-
	{\bf A})^2+E]^{-1} \left((1-\lambda)^{-1}W\right)^{1/2}
$$
is the Birman-Schwinger kernel for the operator
$\widetilde{H}=\mfr1/3[(i\nabla -{\bf A})^2 -3(1-\lambda)^{-1}W]$.
The Birman-Schwinger principle implies that
$\int_0^{\infty} N_1(\widetilde{K}_E) \d E $
is the sum of the negative eigenvalues of  $\widetilde{H}$.
An estimate on this quantity follows from the standard Lieb-Thirring
inequality, i.e.,
$$
	{\int_0^{\infty} N_{1-\lambda}(K_E^>) \d E\leq
	(0.24) \mfr1/3 \int
	\left({3\over (1-\lambda)}W(x)\right)^2\d x
	\leq \mfr3/4 (1-\lambda)^{-2}\int W(x)^2\d x}.
$$
The constant $0.24$ can be found as $L_{1,2}$
in [40], Eq. (51). It was improved slightly by
Blanchard and Stubbe [41], see also [42]. In these references
only the case ${\bf A}=0$ was considered. It is, however, a simple
consequence of the diamagnetic inequality (see [43])
that the constant is independent of ${\bf A}$. \hfill{\bf Q.E.D.}

The Lieb-Thirring inequality in Theorem~5.2 implies  an estimate
on the kinetic energy
$$
	T_\psi = \left\langle \psi \biggl\vert \sum \limits^N_{j=1}
	H^{(j)}_{\rm kin} \biggr\vert \psi \right\rangle
$$
in terms of the one-particle density
$$\uprho_\psi (x) = N\sum\limits_{\sigma_1 = \pm 1/2}
\dots \sum \limits_{\sigma_N = \pm 1/2} \int \limits_{\R^{2(N-1)}} \vert
\psi
(x,x_2 \dots , x_N; \sigma_1, \dots , \sigma_N) \vert^2 \d x_2 \dots \d
x_N.
$$
Here $\psi$ is a normalized $N$ particle fermionic wave function.

{\it 5.3 COROLLARY (Kinetic energy inequality in 2 dimensions).}  {\it
Let
$T_\psi$ and $\uprho_\psi$ be defined as above.
Then for all $0<\lambda<1$ we have
$$T_\psi \geq \cases{0,&if $\uprho_\psi\leq \lambda^{-1}{B \over
\pi}$\cr\cr
		     {1\over 3}(1-\lambda)^2
		     \int\left[\uprho_\psi(x)-\lambda^{-1}{B \over \pi}\right]^2\d x,
			&if $\uprho_\psi\geq \lambda^{-1}{B \over \pi}$}.
\eqno(5.22)$$}

{\it Proof:\/} The inequality in Theorem~5.2 holds
for the operator $H_{\rm kin}-W$ if $|\gamma|\geq1$. If $|\gamma|<1$,
however, one should choose $\Pi_0$ in the proof of Theorem~5.2 as the
projection onto
the levels $\nu=0$ and $\nu=1$ (not only onto $\nu=0$).
Equation (5.21) is then no longer an identity but a bound.
In this way one concludes that the negative eigenvalues
$e_1(W),e_2(W),\ldots$
for $H_{\rm kin}-W$ satisfy
$$
        \sum_j |e_j(W)|\leq  \alpha \int_{\R^2} |W|_+
(x) \d x + \beta \int_{\R^2} |W|_+^2 (x) \d x,
$$
with $\alpha=\lambda^{-1}{B \over \pi}$ and $\beta=\mfr3/4 (1-\lambda)^{-2}
$.
This bound is clearly valid for all $\gamma$.

The proof of (5.22) now follows by a standard Legendre transformation.
In fact, if $W\geq 0$ we have
$$
	T_\psi=\left\langle\psi\biggl\vert \sum_j H^{(j)}_{\rm kin}-W(x_j)
	\biggr\vert \psi
	\right\rangle+\int W\uprho_\psi\geq\int_{\R^2}
	\left[W\uprho_\psi -\alpha  W-\beta W^2\right]\eqno(5.23)
$$
Since the Legendre transformation of the function $W\mapsto
\alpha W +\beta W^2$ is the function
$$
	\uprho\mapsto\sup_{W\geq0}\left[\uprho W-\alpha W-\beta W^2\right]
	=\cases{0,&if $\uprho\leq\alpha$\cr
	        (4\beta)^{-1}(\uprho-\alpha)^2,&if $\uprho>\alpha$},
$$
we see that (5.22) follows by making the optimal choice for $W$ in (5.23).
\hfill{\bf Q.E.D.}

{\it 5.4 LEMMA (Exchange inequality in 2 dimensions).}  {\it Let $\psi \in
\bigotimes \limits^N L^2 (\R^2; \C^2)$ be any normalized $N$ particle
wave-function (not necessarily fermionic) and let
$$\uprho_\psi (x) = \sum \limits^N_{i=1} \sum \limits_{\sigma_1 = \pm 1/2}
\dots \sum \limits_{\sigma_N = \pm 1/2} \int \limits_{\R^{2N}} \vert \psi
(x_1, \dots , x_N; \sigma_1, \dots , \sigma_N) \vert^2 \d x_1
\dots \d x_{i-1} \d x_{i+1}
\dots \d x_N$$
be the corresponding one-particle density.  Then
$$\eqalignno{\sum \limits_{\sigma_1}
\dots \sum \limits_{\sigma_N} &\int \limits_{\R^{2N}} \vert \psi \vert^2
\sum
\limits_{1 \leq i < j \leq N} \vert x_i - x_j \vert^{-1} \d x_1 \dots
\d x_N \cr
&\geq \mfr1/2 \int\limits_{\R^2} \int \limits_{\R^2} \uprho_\psi (x)
\uprho_\psi (y) \vert x-y \vert^{-1} \d x \d y - 192 (2\pi)^{1/2}\int
\limits_{\R^2}
\uprho_\psi (x)^{3/2} \d x. \qquad&(5.24) \cr}$$}

{\it Proof:}  The proof is essentially the same as in [34], where
the three dimensional equivalent of (1.1) was proved.  Our presentation
is inspired by [44].

We use the representation (in 3 dimensions a similar representation was
originally used by Fefferman and de La Llave [45])
$$\vert x-y \vert^{-1} = \pi^{-1} \int \limits_{\R^2} \int \limits_\R
\upchi_R
(x-z) \upchi_R (y-z) R^{-4} \d R\d z, \eqno(5.25)$$
where $\upchi_R$ is the characteristic function of the ball of radius $R$
centered at the origin.  If we use (5.25) to represent $\sum \limits_{i<j}
\vert x_i - x_j \vert^{-1}$ we can estimate the integrand as follows.
$$\eqalignno{\sum \limits_{1 \leq i < j \leq N} &\upchi_R (x_i -z) \upchi_R
(x_j - z)
= \mfr1/2 \left( \sum \limits_i \upchi_R (x_i - z) \right)^2 - \mfr1/2 \sum
\limits_i \upchi_R (x_i - z) \cr
&= \mfr1/2 \left( \sum \limits_i \upchi_R (x_i - z) - \int
\uprho_\psi (y) \upchi_R
(y-z) \d y \right)^2 + \sum \limits_i \upchi_R (x_i - z) \int \uprho_\psi
(z)
\upchi_R
(y-z) \d y \cr
&- \mfr1/2 \left( \int \uprho_\psi (y) \upchi_R (y-z) \d y \right)^2 -
\mfr1/2
\sum \limits_i \upchi_R (x_i - z) \cr
&\geq \sum \limits_i \upchi_R (x_i - z) \int \uprho_\psi (y)
\upchi_R (y-z) \d y -
\mfr1/2 \left( \int \uprho_\psi (y) \upchi_R (y-z) \d y \right)^2 - \mfr1/2
\sum
\limits_i \upchi_R (x_i - z). \cr}$$
If we integrate this inequality over the measure $R^{-4} \d R \d z$, the
last
term, $\mfr1/2 \sum \limits_i \upchi_R (x_i - z)$, will give a divergent
integral.
For the purpose of a lower bound, however, we can restrict the integration
in (5.25)
to $R > r(z)$, where $r(z) > 0$ is some specific function we shall choose
below.  Using the fact that
$$\sum \limits_{\sigma_1} \dots \sum \limits_{\sigma_N} \int
\limits_{\R^{2N}}
\vert \psi \vert^2 \sum \limits_i \upchi_R (x_i - z) \d x_1 \dots \d x_N =
\int
\limits_{\R^2} \uprho_\psi (y) \upchi_R (y-z) \d y,$$
we obtain,
$$\eqalignno{&\sum \limits_{\sigma_1} \dots \sum \limits_{\sigma_N} \int
\limits_{\R^{2N}} \vert \psi \vert^2 \sum \limits_{i < j} \vert x_i - x_j
\vert^{-1} \d x_1 \dots \d x_N \cr &\geq \mfr1/2 \pi^{-1} \int \limits_{\R^2}
\int_{R > r (z)} \left( \int \uprho_\psi (y) \upchi_R (y-z) \d y \right)^2
R^{-4} \d R \d z \cr
&- \mfr1/2 \pi^{-1} \int \limits_{\R^2} \int_{R > r(z)} \int \uprho_\psi
(y) \upchi_R (y-z) \d y R^{-4} \d R \d z \cr
&\geq \mfr1/2 \pi^{-1} \int\int \uprho_\psi (x) \uprho_\psi (y) \vert x-y
\vert^{-1} \d x \d y \cr
&- \mfr1/2 \pi^{-1} \int \limits_{\R^2} \int_{R < r(z)} \pi^2 \uprho^*_\psi
(z)^2 \d R \d z - \mfr1/2 \pi^{-1} \int \limits_{\R^2} \int_{R > r(z)}
R^{-2}
\pi \uprho^*_\psi (z) \d R \d z. \qquad&(5.26)\cr}$$
Here we have introduced the Hardy-Littlewood maximal function
$$\uprho^*_\psi (z) = \sup \limits_R (\pi R^2)^{-1} \int \uprho_\psi (y)
\upchi_R (y-z) \d y, $$
which, viewed as a map from $L^p (\R^2)$ to $L^p (\R^2)$, is a bounded map
for all $p > 1$ (see [46], pp. 54-58).
The error terms in (5.26) can be computed as
$$\eqalignno{\mfr1/2 \pi^{-1} &\int \limits_{\R^2} \left[ \int_{R< r(z)}
\pi^2 \uprho^*_\psi (z)^2 \d R + \int_{R > r(z)} R^{-2} \pi \uprho^*_\psi
(z)
\d R \right] \d z \cr
&= \mfr1/2 \pi^{-1} \int \limits_{\R^2} (r(z) \pi^2 \uprho^*_\psi (z)^2 +
r(z)^{-1} \pi \uprho^*_\psi (z)) \d z. \cr}$$
The optimal choice for $r(z)$ is $r(z) = (\pi \uprho^* (z))^{-1/2}$.  This
means that the error is $\pi^{1/2} \int \limits_{\R^2} \uprho^*_\psi
(z)^{3/2} \d z$, but this can be estimated by the maximal inequality to
be less than $192 (2\pi)^{1/2} \int \limits_{\R^2} \uprho (z)^{3/2} \d z$.
\hfill {\bf Q.E.D.}

\bigskip\noindent
{\bf 5.5  THE LOWER BOUND}

Our goal here is to give a lower bound to $E^{\rm Q} (N,B,V)$ in terms of
$E^{\MTF} (N,B,V)$ with errors of lower order than $N^2$ as $N$ tends to
infinity.  It is important here that $V = N v$, where
$v$ is fixed.  To be more precise we shall prove that
$$N^{-2} [E^{\rm Q} (N,B,V) - E^{\MTF} (N,B,V)] \geq - \varepsilon_N^- (v),
\eqno(5.27)$$
where $\varepsilon_N^- (v)$ is a non-negative function which tends to 0 as
$N \rightarrow \infty$ for fixed $v$.  Note, however, that
$\varepsilon_N^- (v)$ does not depend on $B$.

We shall treat the cases of large $B$ and small $B$ separately.  In the
large $B$ regime we prove (5.27) by a comparison with the classical models
discussed in Sects.~III and IV.  In the small $B$ regime we use magnetic
coherent states, Theorem 5.2 and Lemma 5.4.

The dividing line between large and small $B$ is determined as follows.  If
the minimizer $\uprho^{\rm C}_{1,v}$ of $\E^{\rm C},$ with $\int
\uprho^{\rm C}_{1,v} = 1$ and
confining potential $v$, is bounded
(e.g. for $v(x) = \vert x\vert^2$) then we define small $B$ to mean
$$B/N \leq \beta_c := 2 \pi \sup \limits_x \uprho^{\rm C}_{1,v} (x)
\eqno(5.28)$$

As explained in (3.20) we have for $\beta\geq\beta_c$ that
$E^{\MTF}({1,\beta,v})=E^{\rm C}(1,v)$.

For the general class of $v$ where we do not know that the minimizer
$\uprho^{\rm C}_{1,v}$ is bounded
we  simply define
$$
	\beta_c=N^{1/3}.
$$
By Theorem~3.4 we then have that the function
$$	\delta(N,v)=\sup_{\beta\geq\beta_c}
	|E^{\MTF}({1,\beta,v})-E^{\rm C}(1,v)|\eqno(5.29)$$
tends to zero as $N$ tends to infinity.

{\it Case 1, $B/N \geq \beta_c$:} By simply ignoring the kinetic energy
operator, which we had normalized to be positive, we have the obvious
inequality $E^{\rm Q} (N,B,V) \geq E^{\rm P} (N,V)$ where $E^{\rm P}$ is
the energy of the classical point problem.

{F}rom Lemma 4.3 we can therefore conclude that $$E^{\rm Q} (N,B,V) \geq
E^{\rm P} (N,V) \geq E^{\rm C} (N,V) - b(v) N^{3/2}. \eqno(5.30)$$ Since
$E^{\rm C} (N,V) = N^2 E^{\rm C}(1,v)$ and
$E^{\MTF}(N,B,V)=N^2E^{\MTF}(1,B/N,v)$ we have from (5.29) that

$$E^{\rm Q} (N,B,V) \geq E^{\MTF} (N,B,V) - \delta(N,v) N^2 - b(v) N^{3/2}.
\eqno(5.31)$$
Thus (5.27) holds with $\varepsilon_N^-(v) = \delta(N,v) + b(v) N^{-1/2}$.

We emphasize again that if $\uprho^{\rm C}_{1,v}$ is bounded (e.g.~for
$v=k\vert x\vert^2$)
then $\delta(N,v)$ is not needed.


{\it Case 2, $B/N \leq \beta_c$:}  In this case we use inequality (5.24) to
reduce the many-body problem to a one-body problem.

Let $\psi$ be the many-body ground state\footnote{$^*$}{\ninepoint Since
the exterior potential $V$ tends to infinity at infinity $H_N$ will have a
ground state.} for $H_N$. The correlation estimate (5.24) gives $$E^{\rm Q}
(N,B,V) = \langle \psi \vert H_N \vert \psi \rangle \geq \sum
\limits^N_{j=1} \langle \psi \vert H^{(j)}_{\rm kin} + V(x_j) \vert \psi
\rangle + D (\uprho_\psi, \uprho_\psi) - C \int \limits_{\R^2}
\uprho^{3/2}_\psi (x) \d x. \eqno(5.32)$$

We first estimate the last term in (5.32) in terms of the
kinetic energy $T_\psi = \left\langle \psi \biggl\vert \sum \limits^N_{j=1}
H^{(j)}_{\rm kin} \biggr\vert \psi \right\rangle$ of $\psi$. According to
(5.22) we have $$\eqalignno{\int \uprho^{3/2}_\psi&\leq
(\const.)B^{1/2}\hskip-20pt
\int\limits_{\uprho_\psi<(\const.)B}\hskip-15pt\uprho_\psi
+(\const.)\biggl(\int\uprho_\psi\biggr)^{1/2}
\biggl(\,\,\int\limits_{\uprho_\psi\geq(\const.)B}
\hskip-15pt\uprho_{\psi}^2\biggr)^{1/2}
\cr\cr &\leq (\const.) (\sqrt{\beta_c} N^{3/2} + \sqrt{T_\psi}
N^{1/2}).&(5.33)} $$ Hence, for all $0 < \varepsilon < 1$
(we shall later choose $\varepsilon\sim N^{-1/2}$) we have $$E^{\rm
Q} (N,B,V) \geq \mathop{{\sum}^N}_{j=1} \langle \psi \vert (1 -
\varepsilon) H^{(j)}_{\rm kin} + V(x_j) \vert \psi \rangle + D
(\uprho_\psi, \uprho_\psi) - (\const.) (\sqrt{\beta_c} N^{3/2} +
\varepsilon^{-1} N). \eqno(5.34)$$ where we have used that $$\varepsilon
T_\psi - (\const.) \sqrt{T_\psi} N^{1/2} \geq - (\const.) \varepsilon^{-1}
N. \eqno(5.35)$$

To relate (5.34) to the MTF problem we use the inequality
$$0 \leq D (\uprho_\psi - \uprho^{\MTF}, \uprho_\psi -
\uprho^{\MTF}) = - \left\langle \psi \biggl\vert \sum \limits^N_{j=1}
\uprho^{\MTF} * \vert x_j \vert^{-1} \biggr\vert \psi \right\rangle
+ D(\uprho_\psi, \uprho_\psi) + D (\uprho^{\MTF}, \uprho^{\MTF}),
\eqno(5.36)$$
which is a consequence of the positive definiteness of the kernel $\vert
x-y \vert^{-1}$.  Inserting this in (5.34) gives
$$\eqalignno{E^{\rm Q} (N,B,V) &\geq \sum \limits^N_{j=1} \langle \psi
\vert (1 -\varepsilon) H^{(j)}_{\rm kin} + V (x_j) + \uprho^{\MTF} * \vert
x_j
\vert^{-1}
\vert \psi \rangle \cr
&-D (\uprho^{\MTF}, \uprho^{\MTF}) - (\const.) (\sqrt{\beta_c} N^{3/2} +
\varepsilon^{-1} N). \qquad&(5.37)\cr}$$
Since we have normalized the potential to be positive we have that $(1-
\varepsilon)^{-1} V(x) \geq V(x)$ and also $(1 - \varepsilon)^{-1}
\uprho^{\MTF} *\vert x \vert^{-1} \geq \uprho^{\MTF} * \vert x \vert^{-1}$.

Hence
$$\eqalignno{E^{\rm Q} (N,B,V) &\geq (1 - \varepsilon) \left\langle \psi
\biggl\vert \sum \limits^N_{j=1} (H^{(j)}_{\rm kin} + V(x_j) +
\uprho^{\MTF} *
\vert x_j \vert^{-1}) \biggr\vert \psi \right\rangle \cr
&-D (\uprho^{\MTF}, \uprho^{\MTF}) - (\const.) (\sqrt{\beta_c} N^{3/2} +
\varepsilon^{-1} N). \qquad&(5.38)\cr}$$
Obviously
$$\left\langle \psi \biggl\vert \mathop{{\sum}^N}_{j=1}
\left (H^{(j)}_{\rm kin} +
V(x_j)\right) + \uprho^{\MTF} * \vert x_j \vert^{-1} \biggr\vert
\psi \right\rangle \geq
\sum \limits^N_{j=1} e_j, \eqno(5.39)$$
where $e_1, e_2, \dots , e_N$ are the $N$ lowest eigenvalues of the
one-particle Hamiltonian
$$H^{\MTF}_1 = H_{\rm kin} + V(x) + \uprho^{\MTF} * \vert x \vert^{-1} =
H_{\rm kin} + V^{\MTF} (x) \eqno(5.40)$$

We shall estimate $\mathop{{\sum}^N}_{j=1} e_j$ by a straightforward
coherent
states analysis.

Let $f_1, \dots , f_N$ be the $N$ lowest normalized eigenfunctions of
$H^{\MTF}_1$.  For
technical reasons we introduce a modified operator $\widetilde H^{\MTF}_1$
which is obtained from $H^{\MTF}_1$ by replacing $V^{\MTF}$ by the
truncated
potential
$$\widetilde V^{\MTF} (x) = \cases{V^{\MTF} (x) &$\vert x \vert \leq R_v$
\cr
CN &$\vert x \vert \geq R_v$\cr},$$
where $R_v$ is the finite radius given in the appendix and $C = \inf
\nolimits_{\vert x \vert > R_v} V^{\MTF} (x)/N$ is independent of $N$ by
the
scaling (2.9) of MTF theory.  Note that $V^{\MTF} \geq \widetilde
V^{\MTF}$.  Then from (5.11,12) we have
$$\eqalignno{\sum \limits^N_{j=1} e_j &= \mathop{{\sum}^N}_{j=1} \langle
f_j
\vert H^{\MTF}_1 \vert f_j \rangle \geq \sum \limits^N_{j=1} \langle f_j
\vert \widetilde H^{\MTF}_1 \vert f_j \rangle \cr
&= \sum \limits_\nu \int \d u (\varepsilon_\nu (B) + \widetilde V^{\MTF}
(u))
\sum \limits^N_{j=1} \langle f_j \vert \Pi_{\nu u} \vert f_j \rangle \cr
&-r^{-2} N \int (\nabla g)^2 + \sum \limits^N_{j=1} \langle f_j \vert
\widetilde V^{\MTF} - \widetilde V^{\MTF} * g^2_r \vert f_j \rangle .
\qquad&(5.41)\cr}$$
We first consider the last term.  Writing $\mathop{{\sum}^N}\limits_{j=1}
\vert
f_j (x) \vert^2 = \widetilde{\uprho} (x)$ we have
$$\eqalignno{\mathop{{\sum}^N}_{j=1} &\langle f_j \vert \widetilde V^{\MTF}
-
\widetilde V^{\MTF} * g^2_r (x) \vert f_j \rangle = \int \limits_{\vert x
\vert
< R_v + r} [\widetilde V^{\MTF} (x) - \widetilde V^{\MTF} * g^2_r (x)]
\widetilde{\uprho} (x) \d x \cr
&\geq \int \limits_{\vert x \vert \leq R_v -r} [V^{\MTF} (x) - V^{\MTF}
* g^2_r (x)] \widetilde{\uprho} (x) \d x - \int \limits_{R_v - r \leq \vert
x
\vert \leq R_v + r} \widetilde V^{\MTF} * g^2_r (x) \widetilde{\uprho} (x)
\d x. \cr}$$
Since $V^{\MTF} = N v^{\MTF}$ and $\int \widetilde{\uprho} = N$ we have
$$\mathop{{\sum}^N}_{j=1} \langle f_j \vert \widetilde V^{\MTF} -
\widetilde
V^{\MTF} * g^2_r \vert f_j \rangle \geq - N^2 \sup \limits_{\vert x \vert <
R_v} \vert v^{\MTF} (x) - v^{\MTF} * g^2_r (x) \vert -
\bigl(\sup\limits_{\vert x
\vert < R_v + r} v^{\MTF} (x) \bigr) R^2_v r N^2.
\eqno(5.42)$$
We can then write (5.41) as
$$\sum \limits^N_{j=1} e_j \geq \sum \limits_\nu \int (\varepsilon_\nu
(B) + \widetilde V^{\MTF} (u)) \mathop{{\sum}^N}_{j=1} \langle f_j \vert
\Pi_{\nu u} \vert f_j \rangle \d u - N^2 \widetilde \varepsilon_N (v),
\eqno(5.43)$$
where
$$\eqalignno{\widetilde \varepsilon_N (v) &= \sup \limits_{\vert x \vert <
R_v} \vert v^{\MTF} (x) - v^{\MTF} * g^2_r (x) \vert \cr
&+ \bigl( \sup \limits_{\vert x \vert < R_v + r} v^{\MTF} (x) \bigr)
R^2_v r - N^{-1} r^2 \int (\nabla g)^2 \cr
&\leq C(v) N^{-1/3}. \qquad&(5.44)\cr}$$
For the last step we made the choice $r \sim N^{-1/3}$.

We focus next on the first term in (5.43).  It has the form
$$\int (\varepsilon_\nu (B) + \widetilde V^{\MTF} (u))
\uprho_\nu (u) \d u, \eqno(5.45)$$
where we have denoted $\sum \limits^N_{j=1} \langle f_j \vert \Pi_{\nu u}
\vert f_j \rangle$ by $\uprho_\nu (u)$.  These functions satisfy
$$0 \leq \uprho_\nu (u) \leq \Tr \Pi_{\nu u} = d_\nu (B) \eqno(5.46)$$
and
$$\sum \nolimits_\nu \int \uprho_\nu (u) \d u = N. \eqno(5.47)$$
We obtain a lower bound to (5.45) by minimizing over all functions
$\uprho_\nu$ satisfying (5.46) and (5.47).

Minimizers $\uprho_\nu$ can be constructed as follows.  There is a $\mu >
0$ such that
$$\uprho_\nu (u) = \cases{d_\nu (B) &if $\varepsilon_\nu (B) + \widetilde
V^{\MTF} (u) < \mu$ \cr
0 &if $\varepsilon_\nu (B) + \widetilde V^{\MTF} (u) > \mu$ \cr
\leq d_\nu (B) &if $\varepsilon_\nu (B) + \widetilde V^{\MTF} (u) = \mu$
\cr} \eqno(5.48)$$
All families $\uprho_\nu$ satisfying
(5.48) and the constraint (5.47) are minimizers.
Note that it is possible that $\varepsilon_\nu (B) + \widetilde V^{\MTF}
(u)
= \mu$ on an open set of $u$ values.  The minimizers are therefore not
necessarily unique.  The chemical potential
$\mu$ is uniquely determined by (5.48) and the condition (5.47).

We shall now prove that $\mu = \mu^{\MTF}$.  All we have to show is that we
can find functions $\uprho_\nu$ satisfying (5.47) and (5.48) with $\mu =
\mu^{\MTF}$.

We know from the MTF equation, Theorem 2.2, that if $\uprho^{\MTF} (u) = 0$
then $V^{\MTF}
(u) \geq \mu^{\MTF}$.  Since $\widetilde V^{\MTF}$ differs from $V^{\MTF}$
only
on the set $\uprho^{\MTF} = 0$ we may in (5.48) when $\mu = \mu^{\MTF}$
replace $\widetilde V^{\MTF}$ by $V^{\MTF}$.  We then know from the MTF
equation that there are {\it unique} functions $\uprho_\nu$ satisfying
(5.48)
with $\mu = \mu^{\MTF}$ and $\sum \limits_\nu \uprho_\nu (u) =
\uprho^{\MTF}
(u)$.  In fact, in terms of the filling factors (5.17) we have $\uprho_\nu
(u) = f_\nu (u) d_\nu (B)$.
Since $\int \uprho^{\MTF} (u) \d u = N$ we have produced the
functions $\uprho_\nu$ allowing us to conclude that $\mu$ is indeed equal
to $\mu^{\MTF}$.  If we insert these functions in (5.45) we obtain
$$\sum \limits_\nu \int  \varepsilon_\nu (B) \uprho_\nu (u) \d u + \int
V^{\MTF} (u) \uprho^{\MTF} (u) \d u= \int (j_B (\uprho^{\MTF} (u)) +
V^{\MTF}
(u) \uprho^{\MTF} (u))\d u, \eqno(5.49)$$ where the identity follows from
(2.3).

We can now combine (5.38), (5.39), (5.43), (5.44) and (5.49) to arrive at
$$\eqalignno{E^{\rm Q} (N,B,V) &\geq (1 - \varepsilon) E^{\MTF} (N,B,V) -
\varepsilon D (\uprho^{\MTF}, \uprho^{\MTF}) \cr
&- (\const.) (\sqrt{\beta_c} N^{3/2} + \varepsilon^{-1} N)
- C(v) N^{5/3} \qquad&(5.50)\cr}$$
Hence, since $D(\uprho^{\MTF}, \uprho^{\MTF}) \leq E^{\MTF} (N,B,V)$ we
have
$$\eqalignno{N^{-2} (E^{\rm Q} (N,B,V) - E^{\MTF} (N,B,V)) &\geq - 2
\varepsilon
E^{\MTF} (1,B/N, v) \cr
&- (\const.) (\sqrt{\beta_c} N^{-1/2} + \varepsilon^{-1} N^{-1} +
c(v) N^{-1/3}). \cr}$$
Note that $E^{\MTF} (1, B/N,v)$ is bounded by a constant depending only on
$v$.  If we choose $\varepsilon \sim N^{-1/2}$ we find
$$N^{-2} (E^{\rm Q} (N,B,V) - E^{\MTF} (N,B,V) \geq - c^{-}(v)
[\sqrt{\beta_c}N^{-1/2} +
N^{-1/3}].$$
This is equivalent to (5.27) with $\varepsilon_N^-(v)=c^{-}(v)
[\sqrt{\beta_c}N^{-1/2} +
N^{-1/3}]$. In the case when $\rho^{\rm C}$ is bounded $\beta_c$ is a
constant,
otherwise we chose it to be $\beta_c=N^{1/3}$. In both cases  will
$\varepsilon_N^-(v)$ tend to zero as $N$ tends to infinity.
This finishes the proof of the lower bound.

We have proved (1.7). The limits in (1.8) and (1.9) follow immediately from
the
corresponding results for $E^{\MTF}$ proved in Sects.~II and III.
\bigskip
\noindent{\bf 5.6 HOMOGENEOUS EXTERIOR POTENTIALS}

Finally, we shall show how to prove the stronger result Theorem~1.3
for homogeneous exterior potentials.

In this case we do not have that the minimizing MTF density is
supported within a fixed ball.
In fact, the density will spread out as the coupling constant becomes small.

We shall prove that given $\varepsilon>0$ and $k_0$ there is an
$N_\varepsilon$ independent
of $B$ such that for $N\geq N_\varepsilon$ and $K/N\leq k_0$
$$
	\left|E^{\rm Q}(N,B,Kv)/E^{\MTF}(N,B,Kv)-1\right|<\varepsilon.\eqno(5.51)
$$

We consider large and small $K$ in very much
the same way as we did for $B$ in the lower bound above.
We shall see below that we can find a $k_c$ (depending
only on $v$ and $\varepsilon$) such that (5.51) holds for
$K/N\leq k_c$.

In the case of large $K$, i.e., $K/N\geq k_c$ (but $K/N\leq k_0$) the proof
of (5.51) is then identical to the proof of Theorem~1.1 given above.

For small $K$ we again consider the upper and lower bounds separately.
We begin with the upper bound. We proceed as in Sect.~5.1.
We define the trial operator as in (5.17) and (5.18)
except that we replace $\rho^{\MTF}$ by $\rho^{\rm C}_{N,Kv}$.
The estimate (5.20) now becomes
$$
	E^{\rm Q} (N,B,Kv) \leq \E^{\MTF}[\rho^{\rm C}_{N,Kv};B,Kv]+
	N r^{-2} \int (\nabla g(x))^2 \d x +
	NK \sup\limits_{\vert x \vert \leq R_K} [v * g^2_r (x) - v(x)].
$$
Here $R_k$ is the radius of the ball containing the support of
$\rho^{\rm C}_{N,KV}$ for $K=kN$.
According to (3.22) $R_k=k^{-1/(s+1)}R_1$, where $R_1$ is the radius
for $k=1$, which depends only on $v$.

Using the homogeneity of $v$ we have
$$
	v * g^2_r (x) - v(x)= |x|^s\int\left[v\left((x-y)/|x|\right)
 	-v(x/|x|)\right]g^2_r(y)\d y\leq c_3(v)r |x|^{s-1}.
$$
Thus from (5c.1) we obtain since $s>1$
$$
	\eqalign{E^{\rm Q} (N,B,Kv)- \E^{\MTF} [\rho^{\rm C};B,Kv] &
	\leq c_4(v)N\left[ r^{-2}
 +K r R^{s-1}\right]\cr&\leq c_5(v)N\left[ r^{-2} +
 K r (K/N)^{-(s-1)/(s+1)}\right]\cr
 &\leq c_5(v) N^{2/3}(K/N)^{4/(3(s+1))},}
$$
with the choice $r=(K/N)^{-2/(3(s+1))}$.
We also know from (3.25) that
$$
	\E^{\MTF} [\rho^{\rm C};B,Kv]\leq
	(K/N)^{2/(s+1)}N^2\int j_0({\uprho}^{\rm C}_{1,v})
	+E^{\rm C}(N,Kv)
$$
Hence, from (3.21) and (3.17) we obtain
$$
	\eqalign{{E^{\rm Q} (N,B,Kv)}&\leq
	\left(1+{c_6(v) N^2\over E^{\rm C}(N,Kv)}
	\left[N^{-1/3}(K/N)^{4/(3(s+1))}+(K/N)^{2/(s+1)}
	\right]\right){E^{\rm C}(N,Kv)}\cr
	&\leq\left(1-c_7(v) \left[N^{-1/3}(K/N)^{1/(3(s+1))}+(K/N)^{1/(s+1)}
        \right]\right){E^{\rm C}(N,Kv)}
}
$$
(recall that ${E^{\rm C}(N,Kv)}$ is negative, hence the minus sign in the
second line).
It therefore follows from Theorem~3.5 that we can find $k_c$ depending
only on $\varepsilon$ (but not on $B$) such that
$E^{\rm Q} (N,B,Kv)/E^{\MTF}(N,B,Kv)\geq 1-\varepsilon$ for $K/N\leq k_c$.

We turn next to the lower bound.
As in Sect.~5.5 (in the case $B/N\geq\beta_c$) we may ignore the
kinetic energy operator, which we had normalized to
be positive. We then have the obvious inequality $E^{\rm Q} (N,B,V)
\geq E^{\rm P} (N,V)$
where $E^{\rm P}$ is the energy of the classical point problem.

We shall use Lemma~4.3 to compare $E^{\rm P} (N,Kv)$ to $E^{\rm C} (N,Kv)$.
We must, however, first discuss the scaling of $E^{\rm P} (N,Kv)$.
It is clear that if $v$ is homogeneous of degree $s$
then

$$
        \E^{\rm P}(x_1,\ldots,x_1;Kv)=
        k^{1/(s+1)}\E^{\rm P}(k^{1/(s+1)}x_1,\ldots,k^{1/(s+1)}x_1;v).
$$
Therefore, $E^{\rm P} (N,kV)=k^{1/(s+1)}E^{\rm P}(N,V)$, i.e., $E^{\rm P}
(N,kV)$
has the same scaling as $E^{\rm C} (N,Kv)$ [see (3.21)].

{F}rom Lemma 4.3 we  thus find that
$$E^{\rm P} (N,Kv) \geq E^{\rm C} (N,Kv) - b(v) (K/N)^{1/(s+1)}N^{3/2}
	\geq E^{\rm C} (N,Kv)(1+ c_8(v)N^{-1/2}).\eqno(5.52)$$

According to Theorem~3.5 we may thus assume that  $k_c$ is such that
$$
	E^{\rm Q} (N,B,Kv)/E^{\MTF}(N,B,Kv)\leq
(1+\varepsilon/2)(1+c_8(v)N^{-1/2})
	\eqno(5.53)
$$
for $K/N\leq k_c$. We can therefore clearly find $N_\varepsilon$ such that the
right
side of (5.53) is less than $1+\varepsilon$ for $N\geq N_\varepsilon$.
\bigskip\bigskip
\noindent{\bf VI. CONCLUSION}\par
We have analyzed the ground state of a two-dimensional gas
of $N$ electrons  interacting with each other via the (three-dimensional)
Coulomb potential and subject to a confining exterior potential
$V(x)=Kv(x)$ where $K$ is an adjustable coupling constant.
The electrons are also subject to a uniform magnetic field $B$ perpendicular
to the two dimensional plane.

We have found the {\it exact} energy and electron density function $\rho(x)$
to leading order in $1/N$, i.e., in the high density limit. This limit
is achieved by letting $K$ be proportional to $N$ as $N\to\infty$,
thus effectively confining the electrons to a fixed region of space,
independent of $N$.

It turns out  that the answer to the problem depends critically on the behavior
of $B$ as $N\to\infty$. There are three regimes.

\item{(i)} If $B/N\to0$, i.e., $N\gg B$ in appropriate units, then normal
(two-dimensional) Thomas-Fermi theory gives the exact description.
Correlations can be ignored to leading order in this high density
situation.
\item{(ii)} If $B/N=$constant, a modified TF theory in which
the ``kinetic energy density'' is changed from $(\const.) \rho^2$  to a certain
$B$-dependent function of $\rho$ (called $j_B(\rho)$) is exact.
\item{(iii)} If $B/N\to\infty$ then the kinetic energy term can be omitted
entirely and a classical continuum electrostatics theory emerges as the
exact theory. This electrostatics problem is mathematically interesting
in its own right and can be solved in closed form for the customary choice
$v(x)=\vert x\vert^2$.
\smallskip
Related to the continuum problem is an electrostatics problem for point
charged particles. Apart from its mathematical interest, it provides a
crucial lower bound to the energy in case (iii). Another technical
point of some interest is the extension of the Lieb-Thirring inequality
to the two-dimensional particles in a magnetic field which involves
dealing with  a continuum of zero energy modes (i.e., the lowest Landau level).
\bigskip
\noindent
{\bf APPENDIX}

Here we prove that the minimizers for our three semi-classical problems can
be sought among densities that vanish outside some finite radius --- for
which we give an upper bound.  This lemma is in an appendix because it
pertains to several sections of the paper.

{\it A.1.  LEMMA (Finite radius of minimizers).}  {\it Consider the
three cases:  (a) The
classical energy; (b) the classical particle energy; (c) the MTF
energy.  Let $V (x)$ be the confining potential.
We assume that $V(x) \rightarrow + \infty$ as $\vert x \vert \rightarrow
\infty$ in the sense that the number $W(R): = \inf \{ V(x) : \vert x \vert
\geq R \}$ tends to $\infty$ as $R \rightarrow \infty$.

Then there is a radius $R_v$, depending only on $v = V/N$
$$\eqalignno{E^{\rm P} (N,V) &= \inf \left\{ \E^{\rm P} (x_1, \dots , x_N) :
\vert x_i
\vert \leq R_v \ {\rm for \ all} \ i \right\} \cr
E^{\rm C} (N,V) &= \inf \left\{ \E^{\rm C} [\mu; V] : \ {\rm support} \ \mu
\subset \{
x:
\vert x \vert \leq R_v \}, \int d \mu = N \right\} \cr
E^{\MTF} (N,B,V) &= \inf \left\{ \E^{\MTF} [\uprho;B, V] : \uprho (x) = 0 \
{\rm for} \
\vert x \vert > R_v, \int \uprho = N \right\}. \qquad&(a.1)\cr}$$
Furthermore, any minimizing particle distribution measure or density
satisfies the conditions given in braces in (a.1)

A choice for $R_v$, which is far from optimal, is any $R$
satisfying the inequality
$${1 \over N} W(R) \geq (2 + \pi^{-1}) + {1 \over N} \langle V \rangle_1,
\eqno(a.2)$$
with $\langle V \rangle_1$ being the average of $V$ in the unit disc:
$$\langle V \rangle_1 = {1 \over \pi} \int_{\vert x \vert < 1} V(x) dx.$$}

{\it Proof:}  {\it Particle case:}  Suppose that $\vert x_1\vert > R_v$.
Then we move particle 1 inside $D$, the unit disc centered at the origin.
The point, $y$, to which we move particle 1 is not known, so we average the
energy over all choices of $y \in D$.  If we show that this average energy
is less than the original energy then we know that there is some point $y
\in D$ such that the energy is lowered.  Thus, we have to show that
$$V (x_1) + \sum \limits^N_{j=2} \vert x_1 - x_j \vert^{-1} > \langle V
\rangle_1 + {1 \over \pi} \sum \limits^N_{j=2} \int_D \vert y-x_j
\vert^{-1} dy.$$
Noting that $\int_D \vert y-x \vert^{-1} dy \leq \int_D \vert y \vert^{-1}
dy = 2 \pi$, by a simple rearrangement inequality, we see that it suffices
to have $W(R_v) > \langle V \rangle_1 + 2N$, which
agrees with (a.2).

{\it The Classical case:}  If $\mu$ is any measure with $\int d \mu = N$,
we define $\mu^+$ to be $\mu$ restricted to the complement of the closed
disc of radius $R_v$ centered at the origin.  Thus $\mu^+ (A) =
\mu (A \cap \{ x: \vert x \vert > R_v \})$.  Similarly, $\mu^-$ is
$\mu$ restricted to the disc, so that $\mu = \mu^+ + \mu^-$.  Assuming that

$\mu^+ \not= 0$, we replace $\mu$ by $\mu_\varepsilon := (1 - \varepsilon)
\mu^+ + \mu^- + \delta \nu$, where $\nu$ is Lebesgue measure restricted to
the unit disc, $D$, and where $\pi \delta = \varepsilon \int d \mu^+$.
Thus $\int d \mu_\varepsilon = N$.  The change in energy, to
$O(\varepsilon)$ as $\varepsilon \downarrow 0$, is easily seen to be
$$\delta \int_D V(x) dx - \varepsilon \int V(x) \mu^+ (dx) + \delta
\int_D \int_{\R^2} \vert x-y \vert^{-1} dx \mu (dy) - \varepsilon
\int \int \vert x-y \vert^{-1} \mu^+ (dx) \mu (dy)$$
$< \pi \delta \langle V \rangle_1 - \varepsilon W (R_v)
\int d \mu^+ + 2 \pi \delta N$, which is negative by (a.1).

{\it The MTF case:}  This is similar to the classical case, but with two
differences:  (i) The measure $\mu$ is replaced by a function $\uprho$ with
$\int \uprho (x) dx = N$ and (ii) a ``kinetic energy'' term $\int j_B
(\uprho (x)) dx$ is added to the energy.  Point (i) only simplifies
matters.  For point (ii) we note the simple fact that $j_B (\uprho)$ is
bounded above by $\pi\uprho^2 /2$ and its derivative, $j^\prime (\uprho)$ is
bounded above by $\pi\uprho$; this is true for all $B$.  Let us assume that $d
\mu^+ := \uprho^+ (x) dx$ is not zero, with $\uprho^+ (x) = \uprho(x)$ for
$\vert x \vert > R_v$ and $\uprho^+ (x) = 0$ otherwise.  The
argument is as before, but now we must take into account the change in
kinetic energy which, to leading order in $\varepsilon$, is
$$\delta \int_D j_B^\prime (\uprho (x)) dx - \varepsilon \int _B^\prime
(\uprho^+ (x)) dx < \delta \pi\int_D \uprho (x) dx \leq \delta N =
{\varepsilon } \int \uprho^+ (x) dx.$$
The total energy change is then negative by (a.1). \hfill{\bf Q.E.D.}
\bigskip
\bigskip\noindent
{\bf Acknowledgements} We thank Vidar
Gudmundsson and Jari Kinaret for valuable discussions and Kristinn
Johnsen for allowing us to present the picture of electronic densities in
quantum dots that he prepared.
\vfill\eject
\bigskip\noindent {\bf REFERENCES}
\medskip

\item{[1]} H. van Houten, C.W.J. Beenakker, and A.A.M. Staring, in:
{\it Single Charge Tunneling},
eds. H. Grabert, J.M. Martinis and M.H. Devoret, Plenum, New York, (1991).
\vskip 0pt

\item{[2]} M.A. Kastner, {\it The single electron transistor}, Rev. Mod.
Phys. {\bf 64}, 849-859 (1992)

\item{[3]}  P.L. McEuen, E.B. Foxman, J. Kinaret, U. Meirav, M.A. Kastner,
N.S. Wingreen and S.J. Wind, {\it Self consistent addition spectrum of a
Coulomb island in the quantum Hall regime}, Phys. Rev. B {\bf 45},
11419--11422 (1992). \vskip 0pt

\item{[4]} R.C. Ashoori, H.L. Stormer, J.S. Weiner, L.N. Pfeiffer, S.J.
Pearton, K.W. Baldwin, and K.W. West {\it Single-Electron Capacitance
Spectroscopy of Discrete Quantum Levels}, Phys. Rev. Lett. {\bf 68},
3088--3091 (1992)

\item{[5]} R.C. Ashoori, H.L. Stormer, J.S. Weiner, L.N. Pfeiffer, K.W.
Baldwin, and K.W. West {\it N-electron Ground State Energies of a
Quantum Dot in Magnetic Field}, Phys. Rev. Lett. {\bf 71}, 613--616 (1993)

\item{[6]} N.C. van der Vaart, M.P. de Ruyter van Steveninck, L.P.
Kouwenhoven, A.T. Johnson, Y.V. Nazarov, and C.J.P.M. Harmans, {\it
Time-Resolved Tunneling of Single Electrons between Landau Levels in a
Quantum Dot}, Phys. Rev. Lett. {\bf 73}, 320--323 (1994)\vskip 0pt

\item{[7]} O. Klein, C. Chamon, D. Tang, D.M. Abush-Magder, X.-G. Wen,
M.A. Kastner, and S.J. Wind, {\it Conductance of an Artificial Atom in
Strong Magnetic Fields}, preprint, (1994)

\item{[8]}  A. Kumar, S.E. Laux and F. Stern, {\it Electron states in a
GaAs quantum dot in a magnetic field}, Phys. Rev. B. {\bf 42}, 5166-5175
(1990).\vskip 0pt

\item{[9]} C.W.J. Beenakker, {\it Theory of Coulomb-blockade
oscillations in the conductance of a quantum dot}, Phys. Rev. B {\bf 44},
1646-1656 (1991)

\item{[10]} V. Shikin, S. Nazin, D. Heitmann, and T. Demel, {\it
Dynamic response of quantum dots}, Phys. Rev. B, {\bf 43}, 11903--11907
(1991).\vskip 0pt

\item{[11]} V. Gudmundsson, R.R. Gerhardts, {\it Self-consistent model of
magnetoplasmons in quantum dots with nearly parabolic confinement
potentials}, Phys. Rev. B {\bf 43}, 12098-12101 (1991)

\item{[12]} A.H. MacDonald, S.R. Eric Yang, and M.D. Johnson, {\it Quantum
Dots in Strong Magnetic Fields: Stability Criteria for the Maximum
Density Droplet}, Aust. J. Phys. {\bf 46}, 345-58 (1993)

\item{[13]}  P.L. McEuen, N.S. Wingreen, E.B. Foxman, J. Kinaret, U.
Meirav,
M.A. Kastner, and S.J. Wind, {\it Coulomb interactions and energy-level
spectrum of a small electron gas}, Physica B {\bf 189},
70--79 (1993).\vskip 0pt

\item{[14]} S.-R. Eric Yang, A.H. MacDonald, and M.D. Johnson,
{\it Addition Spectra of Quantum Dots in Strong Magnetic
Fields}, Phys.Rev. Letters {\bf 71}, 3194-3197 (1993)

\item{[15]} J.M. Kinaret and N.S. Windgreen, {\it Coulomb blockade and
partially transparent tunneling barriers in the quantum Hall regime},
Phys. Rev. B, {\bf 48}, 11113--11119 (1993)\vskip 0pt

\item{[16]} D.~Pfannkuche, V.~Gudmundsson, P.A.~Maksym,
{\it Comparison of a Hartree, a
Hartree-Fock, and an exact treatment of quantum dot helium}, Phys. Rev. B,
{\bf 47}, 2244-2250 (1993).

\item{[17]} D.~Pfannkuche, V.~Gudmundsson, P. Hawrylak, and R.R.
Gerhardts,
{\it Far-Infrared Response of Quantum Dots: From Few Electron Exitations
to Magnetoplasmons}, Solid State Electronics {\bf 37}, 1221-1226 (1994)

\item{[18]} M. Ferconi and G. Vignale, {\it Current density functional
theory
of quantum dots in a magnetic field}, preprint (1994)


\item{[19]} E.H. Lieb, J.P. Solovej and J. Yngvason, {\it Heavy Atoms in
the Strong Magnetic Field of a Neutron Star}, Phys. Rev. Lett. {\bf
69}, 749--752 (1992). \vskip 0pt

\item{[20]} E.H. Lieb, J. P. Solovej and J. Yngvason, {\it Asymptotics of
Heavy atoms in High magnetic Fields: I. Lowest Landau Band Regions},
Commun. Pure. Appl. Math. {\bf 47}, 513-591 (1994)

\item{[21]} E.H. Lieb, J. P. Solovej and J. Yngvason, {\it Asymptotics of
Heavy atoms in High magnetic Fields: II. Semiclassical Regions}, Commun.
Math. Phys {\bf 161}, 77-124 (1994)

\item{[22]} E. H. Lieb, J.P. Solovej and J. Yngvason, {\it Quantum Dots},
in: Proceedings of the
conference on partial differential equations and mathematical physics,
Birmingham, Alabama, 1994, International Press (in press)

\item{[23]} Tomishima and K. Yonei, {\it Thomas-Fermi Theory for Atoms
in a Strong Magnetic Field}, Progr. Theor. Phys. {\bf 59}, 683--696
(1978).  \vskip 0pt

\item{[24]} I. Fushiki, E.H. Gudmundsson, C. J. Pethick, and J.
Yngvason, {\it Matter in a Magnetic Field in the Thomas-Fermi and
Related Theories}, Ann. Phys. {\bf 216}, 29--72 (1992).  \vskip
0pt

\item{[25]} J. Yngvason, {\it Thomas-Fermi Theory for Matter in a Magnetic
Field as a Limit of Quantum Mechanics}, Lett. Math. Phys. {\bf 22},
107--117 (1991). \vskip 0pt

\item{[26]} E.H. Lieb and H.-T. Yau, {\it The stability and instability of
relativistic matter}, Commun. Math. Phys. {\bf 118}, 177-213 (1988).
\vskip 0pt

\item{[27]}  V. Fock, {\it Bemerkung zur Quantelung des harmonischen
Oszillators in Magnetfeld}, \hfill\break
Z. Phys. {\bf 47}, 446-448 (1928).
\vskip 0pt

\item{[28]} E.H. Lieb, {\it Thomas-Fermi and related theories of atoms
and molecules}, Rev. Mod. Phys. {\bf 53}, 603--641 (1981); {\it
Erratum}, Rev. Mod. Phys. {\bf 54}, 311 (1982). \vskip 0pt

\item{[29]} E.H. Lieb and B. Simon, {\it The Thomas-Fermi Theory of
Atoms, Molecules and Solids}, Adv. in Math. {\bf 23}, 22--116 (1977).
\vskip 0pt

\item{[30]}  E.H. Lieb and M. Loss, unpublished section of a book on
stability of matter.\vskip 0pt

\item{[31]} E.H. Lieb and W. E. Thirring, {\it Bound for the Kinetic Energy
of Fermions Which Proves the Stability of Matter}, Phys. Rev. Lett,
{\bf 35}, 687--689 (1975).  \vskip 0pt

\item{[32]} E.H. Lieb and W.E. Thirring, {\it Inequalities for the moments
of the
eigenvalues of the Schr\"odinger Hamiltonian and their relation to
Sobolev inequalities}, in: {\it Studies in Mathematical Physics: Essays
in Honor of Valentine Bargmann} (E.H. Lieb, B. Simon and A. Wightman
eds.), 269--303, Princeton University Press, 1976. \vskip 0pt

\item{[33]} L. Erd\H os, {\it Magnetic Lieb-Thirring inequalities},
to appear in Comm. Math. Phys.

\item{[34]}  E.H. Lieb, {\it A lower bound for Coulomb energies}, Phys.
Lett. {\bf 70A}, 444-446 (1979).\vskip 0pt

\item{[35]} E.H. Lieb and S. Oxford, {\it An Improved Lower Bound on the
Indirect Coulomb Energy}, Int. J. Quant. Chem. {\bf 19}, 427--439
(1981). \vskip 0pt

\item{[36]} E.H. Lieb, J. P. Solovej, {\it Quantum coherent operators: A
generalization
of coherent states}, Lett. Math. Phys. {\bf 22}, 145--154 (1991).\vskip 0pt

\item{[37]} E.H. Lieb, {\it A Variational Principle for Many-Fermion
Systems}, Phys. Rev. Lett. {\bf 46}, 457--459; Erratum {\bf 47}, 69
(1981). \vskip 0pt

\item{[38]} B. Simon, {\it Functional Integration and Quantum Physics},
Academic Press, 1979

\item{[39]} K. Fan, {\it Maximum Properties and
Inequalities for the Eigenvalues of Completely Continuous Operators},
 Proc.\ Nat.~Acad.~Sci. {\bf 37} (1951)\vskip 0pt

\item{[40]} E.H. Lieb, {\it On Characteristic Exponents in
Turbulence}, Commun. Math. Phys. {\bf 92}, 57-121 (1984). \vskip 0pt

\item{[41]} Ph. Blanchard and J. Stubbe, {\it Phase space bounds for
Schr\"odinger Hamiltonians and applications}, Preprint.\vskip 0pt

\item {[42]} A. Martin, {\it New results on the moments of the eigenvalues
of the
Schr\"odinger Hamiltonian and applications},
Commun.~Math.~Phys. {\bf 129}, 161--168 (1990)

\item{[43]} E.H. Lieb, {\it Kinetic Energy Bounds and their Application
to the Stability of Matter}, in: {\it Schr\"odinger Operators}, Proceedings
of a conference in S{\o}nderborg Denmark 1988, H. Holden and A. Jensen eds.,
Springer Lecture
Notes in Physics {\it 345}, 371-382 (1989).
\vskip 0pt

\item{[44]} V. Bach, {\it Error bound for the Hartree-Fock energy of atoms
and molecules}, Commun.~Math.~Phys. {\bf 147}, 527--548 (1992).

\item{[45]} C. Fefferman and R. de la Llave,
{\it Relativistic stability of matter. I.},
Rev. Iberoamericana 2, 119-215 (1986).

\item {[46]} E.M.~Stein and G.~Weiss, {\it Fourier analysis on Euclidean
spaces}, Princeton University Press, 1971.


\end